\documentclass[epj,nopacs]{svjour}
\usepackage{graphics}
\usepackage{epsfig}
\usepackage{subfig}
\usepackage{amssymb}   % for math 
\begin{document}
\hyphenation{Super-NEMO}
\hyphenation{had-ronic}
\hyphenation{Heidel-berg}
\title{Probing New Physics Models of Neutrinoless Double Beta Decay with SuperNEMO}
\author{R.~Arnold\inst{1}\and C.~Augier\inst{2}\and J.~Baker\inst{3}\and A.S.~Barabash\inst{4}\and A.~Basharina-Freshville\inst{5}\and M.~Bongrand\inst{2}\and V.~Brudanin\inst{6}\and A.J.~Caffrey\inst{3}\and S.~Cebri\'{a}n\inst{7}\and A.~Chapon\inst{8}\and E.~Chauveau\inst{9,10}\and Th.~Dafni\inst{7}\and F.F.~Deppisch\inst{11}\and J.~Diaz\inst{12}\and D.~Durand\inst{8}\and V.~Egorov\inst{6}\and J.J.~Evans\inst{5}\and R.~Flack\inst{5}\and K-I.~Fushima\inst{13}\and I.~Garc\'{i}a~Irastorza\inst{7}\and X.~Garrido\inst{2}\and H.~G\'{o}mez\inst{7}\and B.~Guillon\inst{8}\and A.~Holin\inst{5}\and K.~Holy\inst{14}\and J.J.~Horkley\inst{3}\and Ph.~Hubert\inst{9,10}\and C.~Hugon\inst{9,10}\and F.J.~Iguaz\inst{7}\and N.~Ishihara\inst{15}\and C.M.~Jackson\inst{11}\and S.~Jullian\inst{2}\and M.~Kauer\inst{5}\and O.~Kochetov\inst{6}\and S.I.~Konovalov\inst{4}\and V.~Kovalenko\inst{1,6}\and T.~Lamhamdi\inst{16}\and K.~Lang\inst{17}\and G.~Lutter\inst{9,10}\and G.~Luz\'{o}n\inst{7}\and F.~Mamedov\inst{18}\and Ch.~Marquet\inst{9,10}\and F.~Mauger\inst{8}\and F.~Monrabal\inst{12}\and A.~Nachab\inst{9,10}\and I.~Nasteva\inst{11}\and I.~Nemchenok\inst{6}\and C.H.~Nguyen\inst{9,10}\and M.~Nomachi\inst{19}\and F.~Nova\inst{20}\and H.~Ohsumi\inst{21}\and R.B.~Pahlka\inst{17}\and F.~Perrot\inst{9,10}\and F.~Piquemal\inst{9,10}\and P.P.~Povinec\inst{14}\and B.~Richards\inst{5}\and J.S.~Ricol\inst{9,10}\and C.L.~Riddle\inst{3}\and A.~Rodr\'{i}guez\inst{7}\and R.~Saakyan\inst{5}\and X.~Sarazin\inst{2}\and J.K.~Sedgbeer\inst{22}\and L.~Serra\inst{12}\and Yu.~Shitov\inst{22}\and L.~Simard\inst{2}\and F.~\v{S}imkovic\inst{14}\and S.~S\"{o}ldner-Rembold\inst{11}\and I.~\v{S}tekl\inst{18}\and C.S.~Sutton\inst{23}\and Y.~Tamagawa\inst{24}\and J.~Thomas\inst{5}\and V.~Timkin\inst{6}\and V.~Tretyak\inst{6}\and Vl.I.~Tretyak\inst{25}\and V.I.~Umatov\inst{4}\and I.A.~Vanyushin\inst{4}\and R.~Vasiliev\inst{6}\and V.~Vasiliev\inst{5}\and V.~Vorobel\inst{26}\and D.~Waters\inst{5}\and N.~Yahlali\inst{12}\and A.~\v{Z}ukauskas\inst{26}% etc
}                     % Do not remove
\mail{frank.deppisch@manchester.ac.uk,\\chris.jackson@hep.manchester.ac.uk, soldner@fnal.gov}
\institute{IPHC, Universit\'e de Strasbourg, CNRS/IN2P3, F-67037 Strasbourg, France\and LAL, Universit\'e Paris-Sud 11,  CNRS/IN2P3, F-91405 Orsay, France\and INL, Idaho Falls, Idaho 83415, USA\and Institute of Theoretical and Experimental Physics, 117259 Moscow, Russia\and University College London, WC1E 6BT London, United Kingdom\and Joint Institute for Nuclear Research, 141980 Dubna, Russia\and University of Zaragoza, C/ Pedro Cerbuna 12, 50009 Spain \and LPC Caen, ENSICAEN, Universit\'e de Caen, F-14032 Caen, France\and Universit\'e de Bordeaux, Centre d'Etudes Nucl\'eaires de Bordeaux Gradignan, UMR 5797, F-33175 Gradignan, France\and CNRS/IN2P3, Centre d'Etudes Nucl\'eaires de Bordeaux Gradignan, UMR 5797, F-33175 Gradignan, France\and University of Manchester, M13 9PL Manchester,  United Kingdom\and IFIC, CSIC - Universidad de Valencia, Valencia, Spain\and Tokushima University, 770-8502, Japan \and FMFI, Comenius University, SK-842 48 Bratislava, Slovakia\and KEK,1-1 Oho, Tsukuba, Ibaraki 305-0801 Japan \and USMBA, Fes, Morocco\and University of Texas at Austin, Austin, Texas 78712-0264, USA\and IEAP, Czech Technical University in Prague,  CZ-12800 Prague, Czech Republic\and Osaka University, 1-1 Machikaneyama Toyonaka, Osaka 560-0043, Japan \and Universitat Aut\`onoma de Barcelona, Spain\and Saga University, Saga 840-8502, Japan\and Imperial College London, SW7 2AZ, London, United Kingdom\and MHC, South Hadley, Massachusetts 01075, USA\and Fukui University, 6-10-1 Hakozaki, Higashi-ku, Fukuoka 812-8581 Japan \and INR, MSP 03680 Kyiv, Ukraine\and Charles University, Prague, Czech Republic}
\abstract{
The possibility to probe new physics scenarios of light Majorana neutrino exchange and right-handed currents at the planned next generation neutrinoless double $\beta$ decay experiment SuperNEMO is discussed. Its ability to study different isotopes and track the outgoing electrons provides the means to discriminate different underlying mechanisms for the neutrinoless double $\beta$ decay by measuring the decay half-life and the electron angular and energy distributions.
%
%\PACS{
%      {PACS-key}{discribing text of that key}   \and
%      {PACS-key}{discribing text of that key}
%     } % end of PACS codes
} %end of abstract
\maketitle
%
%------------------------------------------------------------------------------
\section{Introduction}\label{sec:introduction}

\begin{figure*}[!t]
\centering
\subfloat[][]{
\includegraphics[clip,width=0.35\textwidth]{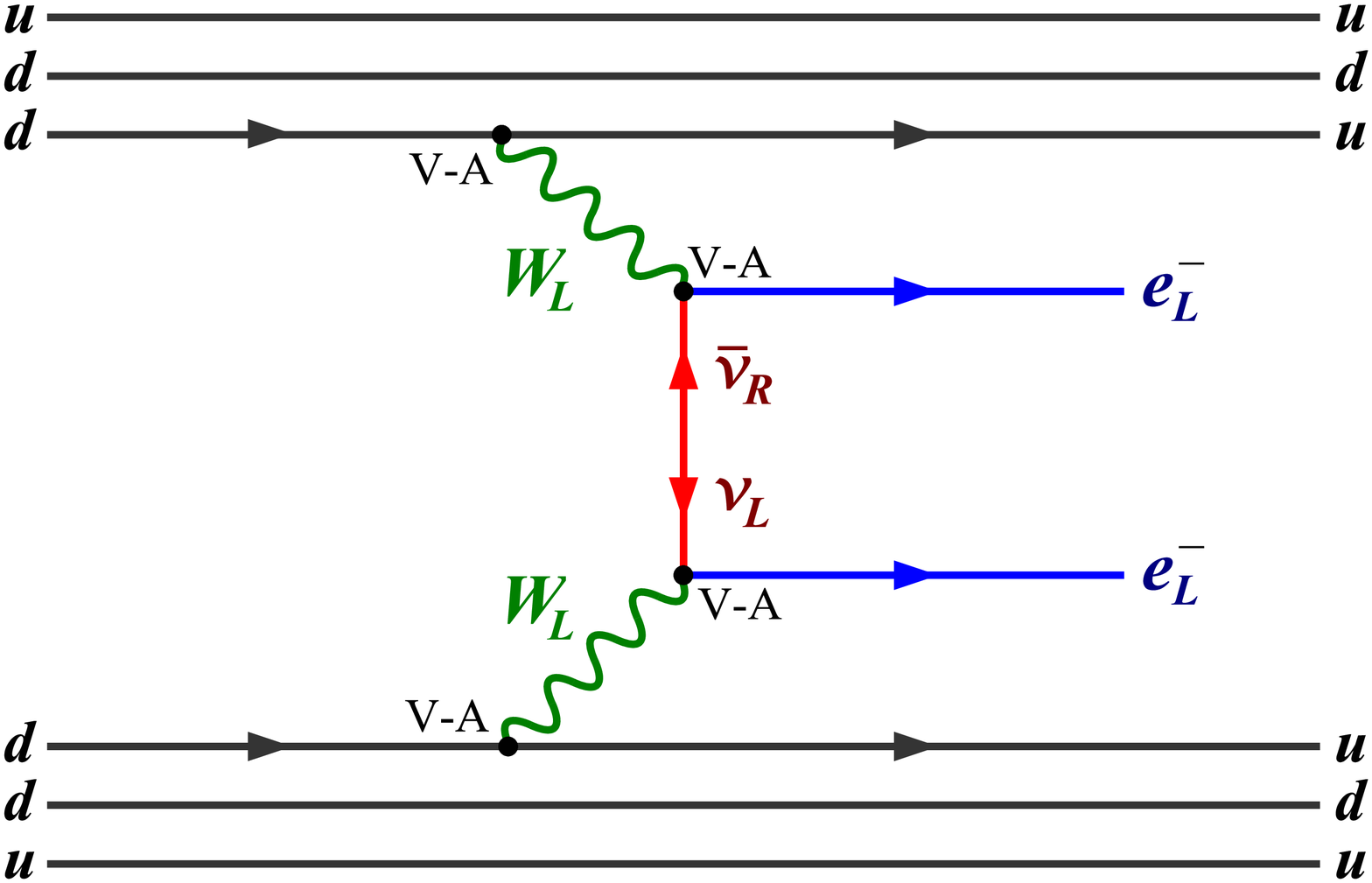}
}
\subfloat[][]{
\includegraphics[clip,width=0.35\textwidth]{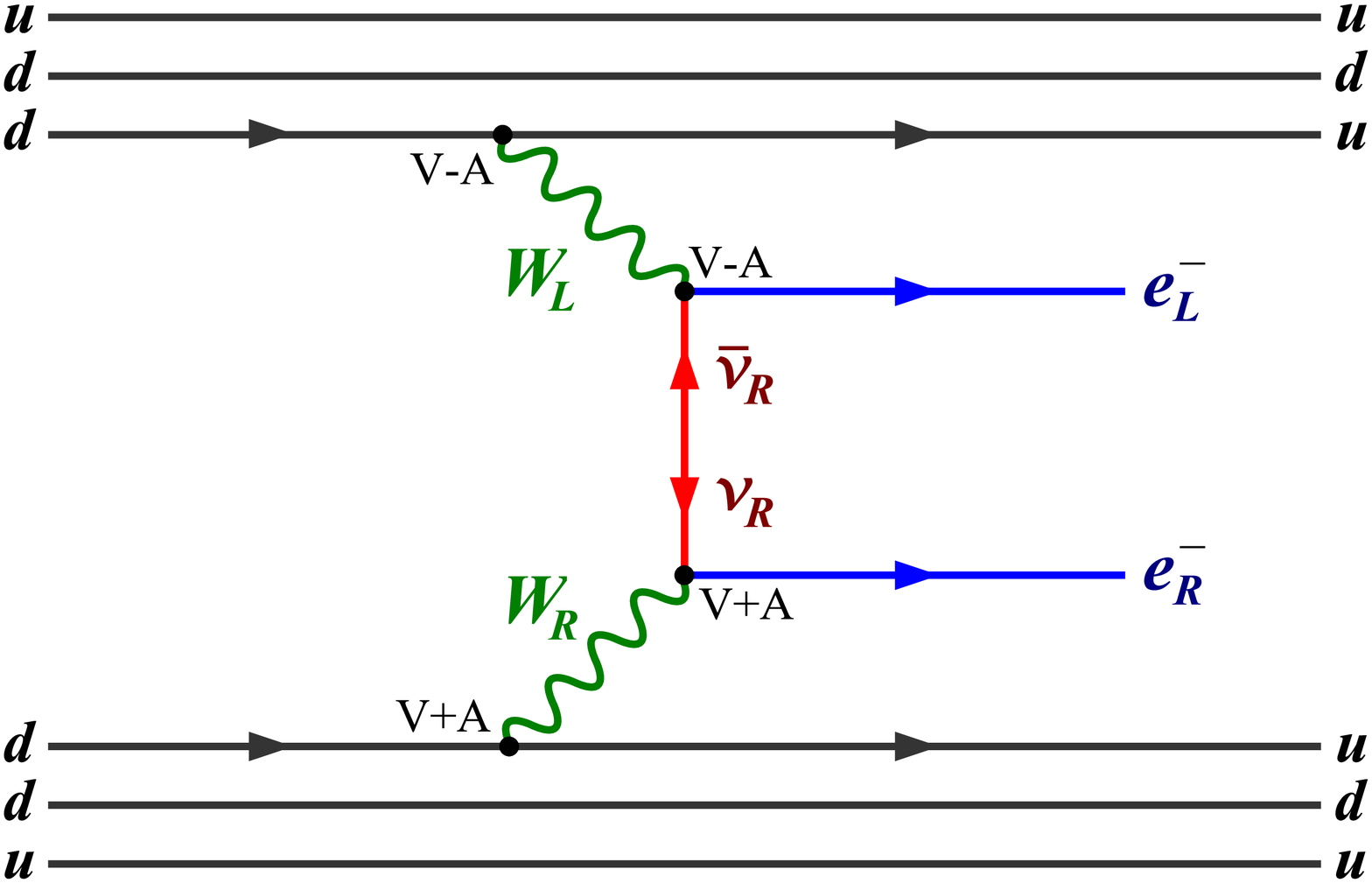}
}
\caption{Diagrams illustrating $0\nu\beta\beta$ decay through (a) the mass mechanism and (b) the right-handed current contribution via the $\lambda$ parameter.} 
\label{fig:0nubbfig}
\end{figure*}

Oscillation experiments have convincingly shown that at least two of the three active neutrinos have a finite mass and that flavour is violated in the leptonic sector~\cite{pdg:2008}. Despite this success, oscillation experiments are unable to determine the absolute magnitude of neutrino masses. Upper limits on the effective electron neutrino mass of 2.3~eV~\cite{kraus} and 2.05~eV~\cite{lobashev} can be set from the analysis of tritium $\beta$ decay experiments. Astronomical observations combined with cosmological considerations yield an upper bound to be set on the sum of the three neutrino masses of the order of 0.7~eV~\cite{Komatsu:2008hk}. However, the most sensitive probe of the absolute mass scale of Majorana neutrinos is neutrinoless double $\beta$ decay ($0\nu\beta\beta$)~\cite{Doi:1982dn,revs,vogel,paesrev}. In this process, an atomic nucleus with $Z$ protons decays into a nucleus with $Z+2$ protons and the same mass number $A$ under the emission of two electrons,
\begin{equation}
	\label{eq:0nubb}
	(A,Z)\to(A,Z+2) + 2 e^-.
\end{equation}
This process can be described by the exchange of a light neutrino connecting two V-A weak interactions, see Fig.~\ref{fig:0nubbfig} (a). The process~(\ref{eq:0nubb}) is lepton number violating and, in the standard picture of light neutrino exchange, it is only possible if the neutrino is identical to its own anti-particle, i.e. if neutrinos are Majorana particles. Combined with the fact that neutrino masses are more than five orders of magnitude smaller than the masses of other fermions, such a possibility suggests that the origin of neutrino masses is different from that of charged fermions. 

Several mechanisms of mass generation have been suggested in the literature, the most prominent example being the seesaw mechanism~\cite{seesaw} in which heavy right-handed neutrinos mix with the left-handed neutrinos and generate light Majorana masses for the observed active neutrinos. The Majorana character of the active neutrinos can then be connected to a breaking of lepton number symmetry close to the GUT scale and might be related to the baryon asymmetry of the Universe through the baryogenesis via leptogenesis mechanism~\cite{leptogenesis}. 

Because of its sensitivity to the nature and magnitude of the neutrino mass, $0\nu\beta\beta$~decay is a crucial experimental probe for physics beyond the Standard Model and its discovery will be of the utmost importance. It will prove lepton number to be broken, and in most models it will also provide direct evidence that the light active neutrinos are Majorana particles\footnote{See~\cite{Bhattacharyya:2002vf} for a counter-example of a model where such a conclusion is not valid.}~\cite{petcov}. However, the measurement of $0\nu\beta\beta$~decay in a single isotope is not sufficient to prove that the standard mechanism of light Majorana neutrino exchange is the dominant source for the decay. There are a host of other models, such as Left-Right symmetry~\cite{Doi:1982dn}, R-parity violating Supersymmetry (SUSY)~\cite{susyacc} or Extra Dimensions~\cite{Bhattacharyya:2002vf}, which can provide alternative mechanisms to trigger $0\nu\beta\beta$~decay. In some of these models, additional sources of lepton number violation can supplement light neutrino exchange. For example, in Left-Right symmetric models, there are additional contributions from right-handed currents and the exchange of heavy neutrinos. In other models, such as R-parity violating SUSY, $0\nu\beta\beta$~decay can be mediated by other heavy particles that are not directly related to neutrinos. 

There are several methods proposed in the literature to disentangle the many possible contributions or at least to determine the class of models that give rise to the dominant mechanism for $0\nu\beta\beta$~decay. Results from $0\nu\beta\beta$~decay can be compared with other neutrino experiments and observations such as tritium decay to determine if they are consistent. At the LHC there could also be signs of new physics exhibiting lepton number violation that is related to $0\nu\beta\beta$ (see~\cite{SUSYLHC} for such an example in R-parity violating Supersymmetry). Such analyses would compare results for $0\nu\beta\beta$ with other experimental searches, but there are also ways to gain more information within the realm of $0\nu\beta\beta$~decay and related nuclear processes. Possible techniques include the analysis of angular and energy correlations between the electrons emitted in the $0\nu\beta\beta$~decay~\cite{Doi:1982dn,Doi:1985,Tomoda:1986,klapdor:2006,Ali:2007ec} or a comparison of results for $0\nu\beta\beta$~decay in two or more isotopes~\cite{Elliot:2004,Bilenky:2004um,Deppisch:2006hb,Gehman:2007qg}. These two approaches are studied in this paper. Other proposed methods are the comparative analysis of $0\nu\beta\beta$~decay to the ground state with either $0\nu\beta^+\beta^+$ or electron capture decay~\cite{Hirsch:1994es} and $0\nu\beta\beta$~decay to excited states \cite{0+}.

Currently, the best limit on $0\nu\beta\beta$~decay comes from the search for  $0\nu\beta\beta$~decay of the isotope $^{76}$Ge giving a half-life of $T_{1/2} > 1.9\cdot10^{25}$~years~\cite{Heidelberg-Moscow}. This results in an upper bound on the $0\nu\beta\beta$ Majorana neutrino mass of $\langle m_{\nu} \rangle\leq 300-600$~meV. A controversial claim of observation of $0\nu\beta\beta$~decay in $^{76}$Ge gives a half-life of $T_{1/2}=(0.8-18.3)\cdot10^{25}$~y~\cite{Klapdor_result} and a resulting effective Majorana neutrino mass of $\langle m_{\nu} \rangle = 110-560$~meV. Next generation experiments such as CUORE, EXO, GERDA, MAJORANA or SuperNEMO aim to increase the half-life exclusion limit by one order of magnitude and confirm or exclude the claimed observation. The planned experiment SuperNEMO allows the measurement of $0\nu\beta\beta$~decay in several isotopes ($^{82}$Se, $^{150}$Nd and $^{48}$Ca are currently being considered) to the ground and excited states, and is able to track the trajectories of the emitted electrons and determine their individual energies. In this respect, the SuperNEMO experiment has a unique potential to disentangle the possible mechanisms for $0\nu\beta\beta$~decay.

This paper addresses the question of how measurements at SuperNEMO can be used to gain information on the underlying physics mechanism of the $0\nu\beta\beta$~decay. The sensitivity of SuperNEMO to new physics parameters in two models is determined by performing a detailed simulation of the SuperNEMO experimental set-up. By analysing both the angular and energy distributions in the standard mass mechanism and in a model incorporating right-handed currents, the prospects of discriminating $0\nu\beta\beta$~decay mechanisms are examined. The two models are specifically chosen to represent all possible mechanisms, as they maximally deviate from each other in their angular and energy distributions. 

This paper is organised as follows. In Section~\ref{sec:theory} a short description of the theoretical framework on which the calculations of the $0\nu\beta\beta$~decay rate and the angular and energy correlations are based is shown. The example physics models are introduced and reviewed. Section~\ref{sec:supernemo} gives a brief overview of the SuperNEMO experiment design and in Section~\ref{sec:simulation} a detailed account of the simulation analysis and its results are presented. In Section~\ref{sec:results}, the expected constraints from SuperNEMO on new physics are shown and the prospects of disentangling $0\nu\beta\beta$ mechanisms by analysing the angular and energy distributions and by comparing rates in different isotopes are addressed. Our conclusions are presented in Section~\ref{sec:conclusion}. 

%------------------------------------------------------------------------------
\section{Neutrinoless Double Beta Decay}\label{sec:theory}

\subsection{Effective Description}\label{sec:EffectiveDescription}
Contributions to $0\nu\beta\beta$ decay can be categorised as either long-range or short-range interactions. In the first case, the corresponding diagram involves two vertices which are both point-like at the Fermi scale, and connected by the exchange of a light neutrino. Such long-range interactions are described by an effective Lagrangian~\cite{Pas:1999fc,Prezeau:2003xn}
\begin{equation}
	\label{eq:longrange}
	{\cal L} = 
	\frac{G_F}{\sqrt{2}}\left(j_{V-A}J_{V-A} + \sum_{a,b}^{L.i.}
	\epsilon^{lr}_{ab} j_{a} J_{b}\right), \label{lrlag}
\end{equation}
where $G_F$ is the Fermi coupling constant and the leptonic and hadronic Lorentz currents are defined as $j_{a} = \bar{e} {\cal O}_a \nu$ and $J_a=\bar{u} {\cal O}_a d$, respectively. Here, ${\cal O}_{a}$ denotes the corresponding transition operator, with $a=V-A,V+A,S-P,S+P,T_L,T_R$~\cite{Pas:1999fc}. In Equation~(\ref{eq:longrange}), the contribution from $V-A$ currents originating from standard weak couplings has been separated off and the summation runs over all Lorentz invariant and non-vanishing combinations of the leptonic and hadronic currents, except for the case $a=b=V-A$. The effective coupling strengths for long-range contributions are denoted as $\epsilon^{lr}_{ab}$.

For short-ranged contributions, the interactions are represented by a single vertex which is point-like at the Fermi scale, and they are described by the Lagrangian~\cite{Prezeau:2003xn,Pas:2000vn}
\begin{eqnarray}\label{lagsr}
	{\cal L} &=& 
	\frac{G^2_F}{2} m_p^{-1} \sum_{a,b,c}^{L.i.} \epsilon^{sr}_{abc} J_a J_b j'_c.
\end{eqnarray}
Here, $m_p$ denotes the proton mass and the leptonic and hadronic currents are given by $J_a=\overline{u}{\cal O}_a d $ and $j'_a=\overline{e}{\cal O}_a e^C$, respectively. The transition operators ${\cal O}_a$ are defined as in the long-range case above, and the summation runs over all Lorentz invariant and non-vanishing combinations of the hadronic and leptonic currents. The effective coupling strengths for the short-range contributions are denoted as $\epsilon^{sr}_{abc}$.

Described by the first term in Equation~(\ref{eq:longrange}), the exchange of light left-handed Majorana neutrinos leads to the $0\nu\beta\beta$ decay rate
\begin{equation}
	\label{eq:t12mm}
	[T_{1/2}^{m_\nu}]^{-1} =
	(\langle m_\nu \rangle/m_e)^2 G_{01}|{\cal M}_{m_\nu}|^2,
\end{equation}
where $\langle m_\nu \rangle$ is the effective Majorana neutrino mass in which the contributions of the individual neutrino masses $m_i$ are weighted by the squared neutrino mixing matrix elements, $U_{ei}^2$, $\langle m_\nu \rangle = |\sum_i U_{ei}^2 m_i|$. 

Analogously, other new physics (NP) contributions, of both long- and short-range nature, can in general be expressed as
\begin{equation}\label{t12np}
	[T_{1/2}^{NP}]^{-1}=\epsilon_{NP}^2 G_{NP} |{\cal M} _{NP}|^2,
\end{equation}
where $\epsilon_{NP}$ denotes the corresponding effective coupling strength, i.e. is either given by $\epsilon^{lr}_{ab}$ for a long-range mechanism or by $\epsilon^{sr}_{abc}$ for a short-range mechanism. In Equations~(\ref{eq:t12mm}) and (\ref{t12np}), the nuclear matrix elements for the mass mechanism and alternative new physics contributions are given by ${\cal M}_{m_\nu}$ and ${\cal M}_{NP}$, respectively, and $G_{01}$, $G_{NP}$ denote the phase space integrals of the corresponding nuclear processes. It is assumed that one mechanism dominates the double $\beta$ decay rate.

\subsection{Left-Right Symmetry}\label{sec:lrsymmetry}
The focus in this paper is on a subset of the Left-Right symmetric model \cite{Doi:1982dn}, which incorporates left-handed and right-handed currents under the exchange of light and heavy neutrinos. Left-Right symmetric models generally predict new gauge bosons of the extra right-handed SU(2) gauge symmetry as well as heavy right-handed neutrinos which give rise to light observable neutrinos via the seesaw mechanism.

\begin{table}[t]
\centering
\begin{tabular}{c|ccc}
Isotope    & $C_{mm}$ [y$^{-1}$] & $C_{\lambda\lambda}$ [y$^{-1}$] & $C_{m\lambda}$ [y$^{-1}$]\\
\hline
$^{76}$Ge  & $1.12\times10^{-13}$ & $1.36\times10^{-13}$ & $-4.11\times10^{-14}$ \\
$^{82}$Se  & $4.33\times10^{-13}$ & $1.01\times10^{-12}$ & $-1.60\times10^{-13}$ \\
$^{150}$Nd & $7.74\times10^{-12}$ & $2.68\times10^{-11}$ & $-3.57\times10^{-12}$ 
\end{tabular}
\caption{Coefficients used in calculating the $0\nu\beta\beta$ decay rate~\cite{Muto:1989cd}.}
\label{tab:NME}
\end{table}

The $0\nu\beta\beta$ decay half-life in the Left-Right symmetric model can be written as a function of the effective parameters $\mu, \eta, \lambda$ \cite{Muto:1989cd},
\begin{eqnarray}
\label{eq:T12_mu_eta_lambda}
\lefteqn{[T_{1/2}]^{-1} =	  
  C_{mm} \mu^2 
   + C_{\lambda\lambda} \lambda^2
   + C_{\eta\eta} \eta^2}
\nonumber \\ 
& &   + C_{m\lambda} \mu\lambda
   + C_{m\eta} \mu\eta
   + C_{\eta\lambda} \eta\lambda,
\end{eqnarray}
where contributions from the exchange of heavy neutrinos are omitted. The coefficients $C_{mm}, C_{\eta\eta}$ etc. are combinations of phase space factors and nuclear matrix elements. The first three terms give the contributions from the following processes:
\begin{enumerate}
   \item $C_{mm}\mu^2$: Fully left-handed current neutrino exchange, see Fig.~\ref{fig:0nubbfig} (a) (mass mechanism);
	\item $C_{\lambda\lambda}\lambda^2$: Right-handed leptonic and right-handed hadronic current neutrino exchange, see Fig.~\ref{fig:0nubbfig} (b);
	\item $C_{\eta\eta}\eta^2$: Right-handed leptonic and left-handed hadronic current neutrino exchange.
\end{enumerate}
The remaining terms in Equation~(\ref{eq:T12_mu_eta_lambda}) describe interference effects between these three processes. The effective parameters $\mu, \eta, \lambda$ in (\ref{eq:T12_mu_eta_lambda}) are given in terms of the underlying physics parameters as
\begin{eqnarray}
\label{eq:muterm}
	\mu     &=& m_e^{-1}   \sum_{i=1}^3 \left(U_{ei}^{11}\right)^2 m_{\nu_i} = \frac{\langle m_\nu\rangle}{m_e}, \\
\label{eq:etaterm}
	\eta    &=& \tan\zeta  \sum_{i=1}^3 U_{ei}^{11}U_{ei}^{12}, \\
\label{eq:lambdaterm}
	\lambda &=& \left(\frac{M_{W_L}}{M_{W_R}}\right)^2  
	            \sum_{i=1}^3 U_{ei}^{11}U_{ei}^{12},
\end{eqnarray}
with the electron mass $m_e$, the left- and right-handed W boson masses $M_{W_L}$ and $M_{W_R}$, respectively, and the mixing angle $\zeta$ between the W bosons. The $3\times 3$ matrices $U^{11}$ and $U^{12}$ connect the weak eigenstates $(\nu_e,\nu_\mu,\nu_\tau)$ of the light neutrinos with the mass eigenstates of the light neutrinos $(\nu_1,\nu_2,\nu_3)$, and heavy neutrinos, $(N_1,N_2,N_3)$, respectively. We assume that the neutrino sector consists of three light neutrino states, $m_{\nu_i} \ll m_e$, and three heavy neutrino states, $M_{N_i} \gg m_p$, $i=1,2,3$. Consequently, the summations in~(\ref{eq:muterm}, \ref{eq:etaterm}, \ref{eq:lambdaterm}) are only over the light neutrino states. For a simple estimate of the sensitivity of $0\nu\beta\beta$ decay to the model parameters, we neglect the flavour structure in $U^{11}$ and $U^{12}$; using the assumption that the elements in $U^{11}$ are of order unity (almost unitary mixing), and those in $U^{12}$ are of order $m_D/M_R \sim \sqrt{m_\nu/M_R}$, with the effective magnitude $m_D$ of the neutrino Dirac mass matrix, and the light and heavy neutrino mass scales, $m_\nu$ and $M_R$, leads to the approximate relations:
\begin{eqnarray}
	\mu     &\approx& \frac{m_\nu}{m_e}, \\
	\eta    &\approx& \tan\zeta\sqrt{\frac{m_\nu}{M_R}}, \\
	\lambda &\approx& \left(\frac{M_{W_L}}{M_{W_R}}\right)^2  
	            \sqrt{\frac{m_\nu}{M_R}}.
\end{eqnarray}
In the following analysis a simplified model incorporating only an admixture of mass mechanism (MM) due to a neutrino mass term $\mu=\langle m_\nu\rangle/m_e$ and right-handed current due to the $\lambda$ term (RHC$_\lambda$) is considered:
\begin{eqnarray}
\label{eq:T12_mu_lambda}
	[T_{1/2}]^{-1} = 
	  C_{mm} \mu^2 
	+ C_{\lambda\lambda} \lambda^2
	+ C_{m\lambda} \mu\lambda.
\end{eqnarray}
As we will see in Section~\ref{sec:distros}, these two mechanisms exhibit maximally different angular and energy distributions, and with an admixture between them, to a good approximation any possible angular and energy distribution can be produced. In our numerical calculation we use the values as given in Table~\ref{tab:NME} for the coefficients $C_{mm}$, $C_{\lambda\lambda}$ and $C_{m\lambda}$ in Equation~(\ref{eq:T12_mu_lambda}). Furthermore, we assume that the parameter $\mu$ is real-valued positive and $\lambda$ is real-valued.

\subsection{Nuclear Matrix Elements}\label{sec:nme}
As demonstrated in Equations~(\ref{eq:t12mm}) and (\ref{t12np}), a calculation of the nuclear matrix elements (NMEs) is required to convert the measured half-life rates or limits into new physics parameters. Exact solutions  for the NMEs do not exist, and approximations have to be used. Calculations using the nuclear shell model exist for lighter nuclei such as $^{76}$Ge and $^{82}$Se, though the only reliable results are for $^{48}$Ca. Quasi-particle random phase approximation (QRPA) calculations are applied for most isotopes as a greater number of intermediate states can be included. In this paper, a comparison between two possible SuperNEMO isotopes ($^{82}$Se and $^{150}$Nd) and the isotope that gives the current best limit ($^{76}$Ge) is made. Consistent calculations of the NMEs for these three isotopes in both the MM and RHC are rare (only~\cite{Muto:1989cd} and~\cite{Tomoda:1991}). All the results are shown using NMEs from~\cite{Muto:1989cd}, displayed in Table~\ref{tab:NME}. 

Recent work on the calculation of NMEs for the heavy isotope $^{150}$Nd suggests that nuclear deformation must be included, as QRPA calculations usually consider the nucleus to be spherical. To compensate for this a suppression factor of 2.7 is introduced into the NME due to an approximation arising from the BCS overlap factor~\cite{deform}, $\mathcal{M}(^{150}$Nd)/2.7. This gives a suppression $C_{mm,\lambda\lambda,m\lambda}/(2.7)^2$ in Table~\ref{tab:NME}. The $^{82}$Se nuclei are assumed to be spherical and no correction is added in this paper.

The NMEs are a significant source of uncertainty in double $\beta$ decay physics and quantitative results in this paper could change with different calculations (particularly for $^{150}$Nd). For example, more recent studies~\cite{Rodin:2007} suggest the NMEs from $^{150}$Nd for the MM are an additional factor 1.3-1.7 smaller. In our analysis we assume a theoretical uncertainty of 30\% in the NMEs of all isotopes and mechanisms considered throughout. Even though the choice of NME changes quantitative results for the extracted physics parameters, the conclusions about the advantages of using different kinematic variables will not be affected.   

\subsection{Angular and Energy Distributions in the Left-Right Symmetric Model}\label{sec:distros}
For our event simulation, the three-dimensional distribution of the $0\nu\beta\beta$ decay rate in terms of the kinetic energies $t_{1,2}$ of the two emitted electrons and the cosine of the angle between the electrons $\cos\theta_{12}$ is used: 
\begin{equation}
        \label{eq:3ddist}
        \rho(t_1,t_2,\cos\theta_{12})=\frac{d\Gamma}{dt_1 dt_2 d\cos\theta_{12}}.
\end{equation}
The distributions for the MM and for the RHC$_\lambda$ mechanism are given by
\begin{eqnarray}
  \label{eq:DistroLHC}
  \lefteqn{\rho_{\mathrm{MM}}(t_1,t_2,\cos\theta_{12}) = } 
  \nonumber \\
  & &	c_{1}\times(t_1+1) p_1 (t_2+1) p_2 F(t_1,Z)F(t_2,Z)\nonumber\\
  & &   \times\delta(Q-t_1-t_2)\left(1-\beta
_1\beta_2\cos\theta_{12}\right),
  \\
  \label{eq:DistroRHC}
  \lefteqn{\rho_{\mathrm{RHC}}(t_1,t_2,\cos\theta_{12}) = }
  \nonumber \\
  & &  c_{2}\times(t_1+1) p_1 (t_2+1) p_2 F(t_1,Z)F(t_2,Z) (t_1-t_2)^2
  \nonumber\\
  & &  \times\delta(Q-t_1-t_2)\left(1+\beta_1\beta_2\cos\theta_{12}\right),
\end{eqnarray}
with the electron momenta $p_i=\sqrt{t_i(t_i+2)}$ and velocities $\beta_i=p_i/(t_i+1)$, and the mass difference $Q$ between the mother and daughter nucleus. All energies and momenta are expressed in units of the electron mass and $c_{1}$ and $c_{2}$ are normalisation constants. The Fermi function $F$ is given by
\begin{equation}
	\label{eq:FermiFunction}
	F(t,Z)=c_{3}\times p^{2s-2} e^{\pi u} |\Gamma(s+i u)|^2,
\end{equation}
where $s=\sqrt{1-(\alpha Z)^2}$, $u=\alpha Z (t+1)/p$, $\alpha=1/137.036$, $\Gamma$ is the Gamma function and $c_3$ is a normalisation constant. Here, $Z$ is the atomic number of the daughter nucleus. The normalisation of the distributions is irrelevant when discussing energy and angular correlations. 

Using Equations~(\ref{eq:DistroLHC}) and (\ref{eq:DistroRHC}), the differential decay widths with respect to the cosine of the angle $\theta_{12}$,
\begin{equation}
	\label{eq:DistroAngle}
	\frac{d\Gamma}{d\cos\theta_{12}} =
	\int_0^Q dt_1 \rho(t_1,Q-t_1,\cos\theta_{12}),
\end{equation}
and the energy difference $\Delta t = t_1-t_2$,
\begin{equation}
	\label{eq:DistroEnergyDiff}
	\frac{d\Gamma}{d(\Delta t)} =
	\int_{-1}^1 d\cos\theta_{12} 
	\rho\left(\frac{Q+\Delta t}{2},\frac{Q-\Delta t}{2},\cos\theta_{12}\right),
\end{equation}
may be determined.

The differential width in Equation (\ref{eq:DistroAngle}) can be written as~\cite{Doi:1982dn,Ali:2007ec}
\begin{equation}\label{eq:angulardistribution}
	\frac{d\Gamma}{d\cos\theta_{12}} =
	\frac{\Gamma}{2}(1-k_{\theta}\cos\theta_{12}),
\end{equation}
with the total decay width \(\Gamma\). The distribution shape is linear in $\cos\theta_{12}$, with the slope determined by the parameter $k_{\theta}$ which can range between \(-1 \leq k_{\theta} \leq 1\), depending on the underlying decay mechanism. Assuming the dominance of one scenario, the shape does not depend on the precise values of new physic parameters (mass scales, coupling constants) but is a model specific signature of the mechanism. For the MM and RHC$_\lambda$ mechanisms, the theoretically predicted $k_{\theta}$ is found from Equation~(\ref{eq:DistroAngle}) and is given by 
\begin{eqnarray}
	k_{\theta_{\mathrm{MM}}}^{\mathrm{Se}} &=& +0.88, \quad k_{\theta_{\mathrm{MM}}}^{\mathrm{Nd}} = +0.89, \\
	k_{\theta_{RHC}}^{Se} &=& -0.79, \quad k_{\theta_{RHC}}^{Nd} = -0.80. 
\end{eqnarray}
The correlation coefficient $k_{\theta}$ is modified when taking into account nuclear physics effects and exhibits only a small dependence on the type of nucleus. The MM and the RHC$_\lambda$ mechanisms give the maximally and minimally possible values for the angular correlation coefficient $k_\theta$ in a given isotope, respectively.

Experimentally, $k_{\theta}$ can be determined via the forward-backward asymmetry of the decay distribution,
\begin{eqnarray}\label{eq:k}
	\lefteqn{\mathcal A_{\theta} \equiv}
\nonumber \\
 & &	\left(\int_{-1}^0 \frac{d\Gamma}{d\cos\theta} d\cos\theta
	- \int_0^1 \frac{d\Gamma}{d\cos\theta} d\cos\theta\right)/\Gamma 
        = 
\nonumber \\
& & \frac{N_{+}-N_{-}}{N_{+}+N_{-}} = \frac{k_{\theta}}{2}. 
\end{eqnarray}
Here, $N_+$ ($N_-$) counts the number of signal events with the angle $\theta_{12}$ larger (smaller) than $90^\circ$.

\begin{figure}[!t]
\centering
\includegraphics[clip,width=0.49\textwidth]{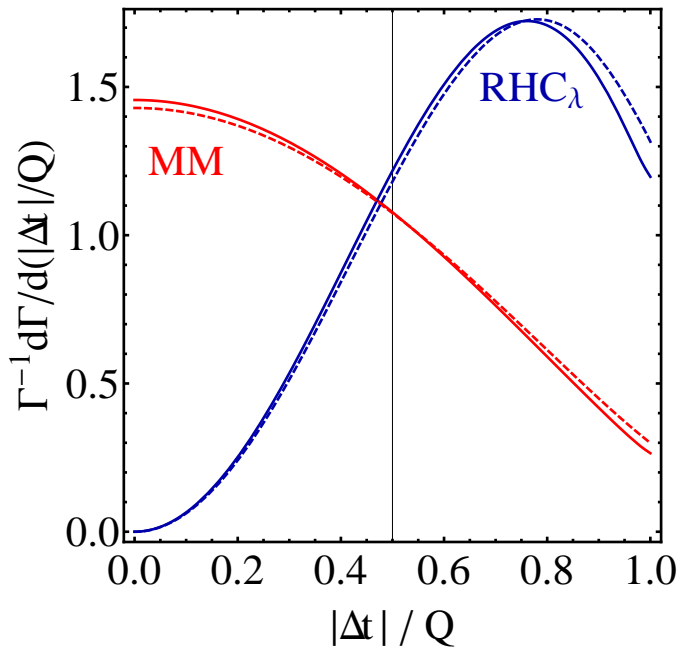}
\caption{Normalised $0\nu\beta\beta$ decay distribution with respect to the electron energy difference in the MM (red) and RHC$_\lambda$ mechanism (blue) for the isotopes $^{82}$Se (solid curves) and $^{150}$Nd (dashed curves).} 
\label{fig:edist}
\end{figure}

Analogously, the MM and RHC$_\lambda$ mechanism also differ in the shape of the electron energy difference distribution, Equation~(\ref{eq:DistroEnergyDiff}). For the isotopes $^{82}$Se and $^{150}$Nd, these distributions are shown in Fig. \ref{fig:edist}.
Again, the shape is largely independent of the isotope under inspection. The following asymmetry in the electron energy-difference distribution is determined,
\begin{eqnarray}\label{eq:kE}
	\lefteqn{\mathcal A_{E} \equiv}
\nonumber \\
& &	\left(\int_{0}^{Q/2} \frac{d\Gamma}{d(\Delta t)} d(\Delta t)
	- \int_{Q/2}^Q \frac{d\Gamma}{d(\Delta t)} d(\Delta t)\right)/\Gamma 
	=
\nonumber \\
& &    \frac{N_{+}-N_{-}}{N_{+}+N_{-}} = \frac{k_E}{2},
\end{eqnarray}
thereby defining an energy correlation coefficient $k_E$, where $Q$ is the energy release of the decay. The rate $N_+$ ($N_-$) counts the number of signal events with an electron energy difference smaller (larger) than $Q/2$.
For the MM and RHC$_\lambda$ mechanism, the theoretical $k_E$ parameter may be found from Equation~(\ref{eq:DistroEnergyDiff}) and is given by 
\begin{eqnarray}
	k_{E_{\mathrm{MM}}}^{\mathrm{Se}} &=& +0.66, \quad k_{E_{\mathrm{MM}}}^{\mathrm{Nd}} = +0.64, \\
	k_{E_{\mathrm{RHC}}}^{\mathrm{Se}} &=& -1.07, \quad k_{E_{\mathrm{RHC}}}^{\mathrm{Nd}} = -1.09,
\end{eqnarray}
in the isotopes $^{82}$Se and $^{150}$Nd. As can be seen in Fig.~\ref{fig:edist}, the MM and RHC$_\lambda$ mechanisms correspond to different shapes of the energy difference distribution. Analogous to the angular distribution, the corresponding energy correlation coefficients in the two mechanisms considered are, to a good approximation, at their upper and lower limits in a given isotope.

%------------------------------------------------------------------------------
%------------------------------------------------------------------------------
\section{SuperNEMO}\label{sec:supernemo}
\begin{figure*}[!t]
\centering
\includegraphics[clip,width=0.37\textwidth]{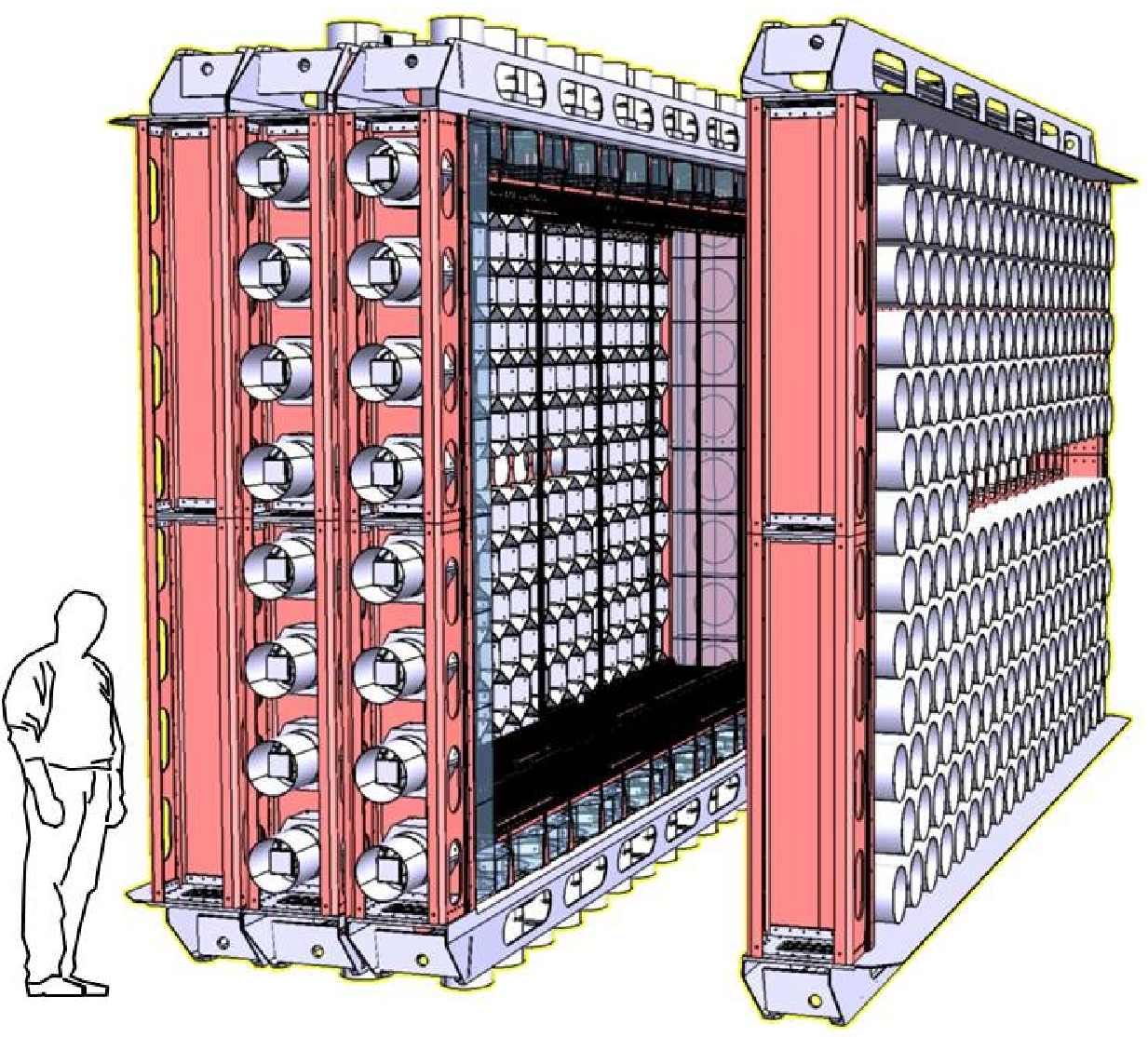}
\label{fig:module-a}
\includegraphics[clip,width=0.37\textwidth]{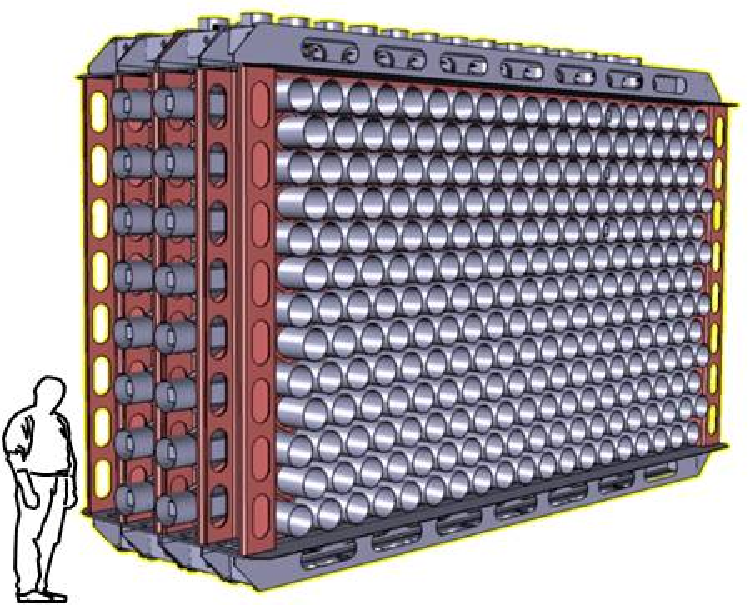}
\label{fig:module-b}
\caption{A SuperNEMO module. The source foil (not shown) sits in the centre of a tracking volume consisting of drift cells operating in Geiger mode. This is surrounded by calorimetry consisting of scintillator blocks connected to PMTs (grey). The support frame is shown in red.} 
\label{fig:module} 
\end{figure*}

SuperNEMO is a next generation experiment building on technology used by the currently running NEMO-III experiment~\cite{nemo3,NEMO_Se82,NEMO_Majarons,NEMO_excitedstates,NEMO_Nd150,NEMO_background,NEMO_latestconf}. The design of the detector consists of 20 modules each containing approximately $5$~kg of enriched and purified double $\beta$ emitting isotope in the form of a thin foil (with a surface density of $40$~mg/cm$^{2}$). Isotopes under consideration for SuperNEMO are $^{82}$Se, $^{150}$Nd and $^{48}$Ca. 

The foil is surrounded by a tracking chamber comprising nine planes of drift cells (44 mm diameter) on each side operating in Geiger mode in a magnetic field of 25 Gauss. The tracking chamber has overall dimensions of 4 m height (parallel to the drift cells), 5 m length and 1 m width (perpendicular to the foil); the foil is centred in this volume with dimensions of 3 m height and 4.5 m length. The tracking allows particle identification $(e^{-}, e^{+}, \gamma, \alpha)$ and vertex reconstruction to improve background rejection and to allow measurement of double $\beta$ decay angular correlations. 

Calorimetry consisting of 25$\times$25~cm$^{2}$ square blocks of 5 cm thickness scintillating material connected to low activity photomultiplier tubes (PMTs) surrounds the detector on four sides. An energy resolution of 7$\%$ (FWHM) and time resolution of 250 ps (Gaussian $\sigma$) at 1 MeV for the blocks is required.  The granularity of the calorimetry allows the energy of individual particles to be measured. Additional $\gamma$-veto calorimetry to identify photons from background events of thickness 10 cm surrounds the detector on all sides. The modules are contained in shared background shielding and will be housed in an underground laboratory to reduce the cosmic ray flux. A diagram of the planned SuperNEMO module design is shown in Fig.~\ref{fig:module}.

%------------------------------------------------------------------------------
\section{Simulation}\label{sec:simulation}
\subsection{Simulation Description}\label{sec:sd}
A full simulation of the SuperNEMO detector was performed including realistic digitisation, tracking and event selection. Signals for two mechanisms of $0\nu\beta\beta$ decay (mass mechanism MM and right-handed current via the $\lambda$ parameter RHC$_\lambda$) and the principal internal backgrounds were generated using DECAY0~\cite{decay0}. This models the full event kinematics, including angular and energy distributions. 

A GEANT-4 Monte Carlo simulation of the detector was constructed. Digitisation of the hits in cells was obtained by assuming a Geiger hit model validated with NEMO-III with a transverse resolution of 0.6 mm and a longitudinal resolution of 0.3 cm. The calorimeter response was simulated assuming a Gaussian energy resolution of 7\%/$\sqrt{E}$ (FWHM) and timing resolution of 250 ps (Gaussian $\sigma$ at 1 MeV). Inactive material in front of the $\gamma$-veto was partially simulated.

Full tracking was developed consisting of pattern recognition and helical track fitting. The pattern recognition uses a cellular automaton to select adjacent hits in the tracking layers. Helical tracks are fitted to the particles. Tracks are extrapolated into the foil to find an appropriate event origin and into the calorimeter where they may be associated with calorimeter energy deposits. The realistic event selection (validated using NEMO-III) was optimised for double $\beta$ decay electrons (two electrons with a common vertex in the foil). The selection criteria are:
\begin{itemize}
\item events must include only two negatively charged particles each associated with one calorimeter hit;
\item event vertices must be within the foil and the tracks must have a common vertex of \textless10 standard deviations between intersection points in the plane of the source foil;
\item the time of flight of the electrons in the detector must be consistent with the hypothesis of the electrons originating in the source foil;
\item the number of Geiger drift cell hits unassociated with a track must be less than 3;
\item the energy deposited in individual calorimeter blocks must be $>50$~keV;
\item there are zero calorimeter hits not associated with a track;
\item tracks must have hits in at least one of the first three and one of the last three planes of Geiger drift cells;
\item the number of delayed Geiger drift cell hits due to $\alpha$ particles from $^{214}$Bi-$^{214}$Po events must be zero;
\item there are no hits in the $\gamma$-veto detectors with energy $>50$~keV.
\end{itemize}

Using these experimental selection criteria the signal efficiency was found to be 28.2\% for the MM and 17.0\% for the RHC$_{\lambda}$ in $^{82}$Se and 29.1\% for the MM and 17.3\% for the RHC$_{\lambda}$ in $^{150}$Nd. This is higher than the efficiency for MM detection in $^{100}$Mo decays in NEMO-III of 17.4\% (in the electron energy sum range 2-3.2 MeV)~\cite{NEMO_latestconf}. 

\begin{figure*}[!t]
\centering
\subfloat[][]{
\includegraphics[clip,width=0.37\textwidth]{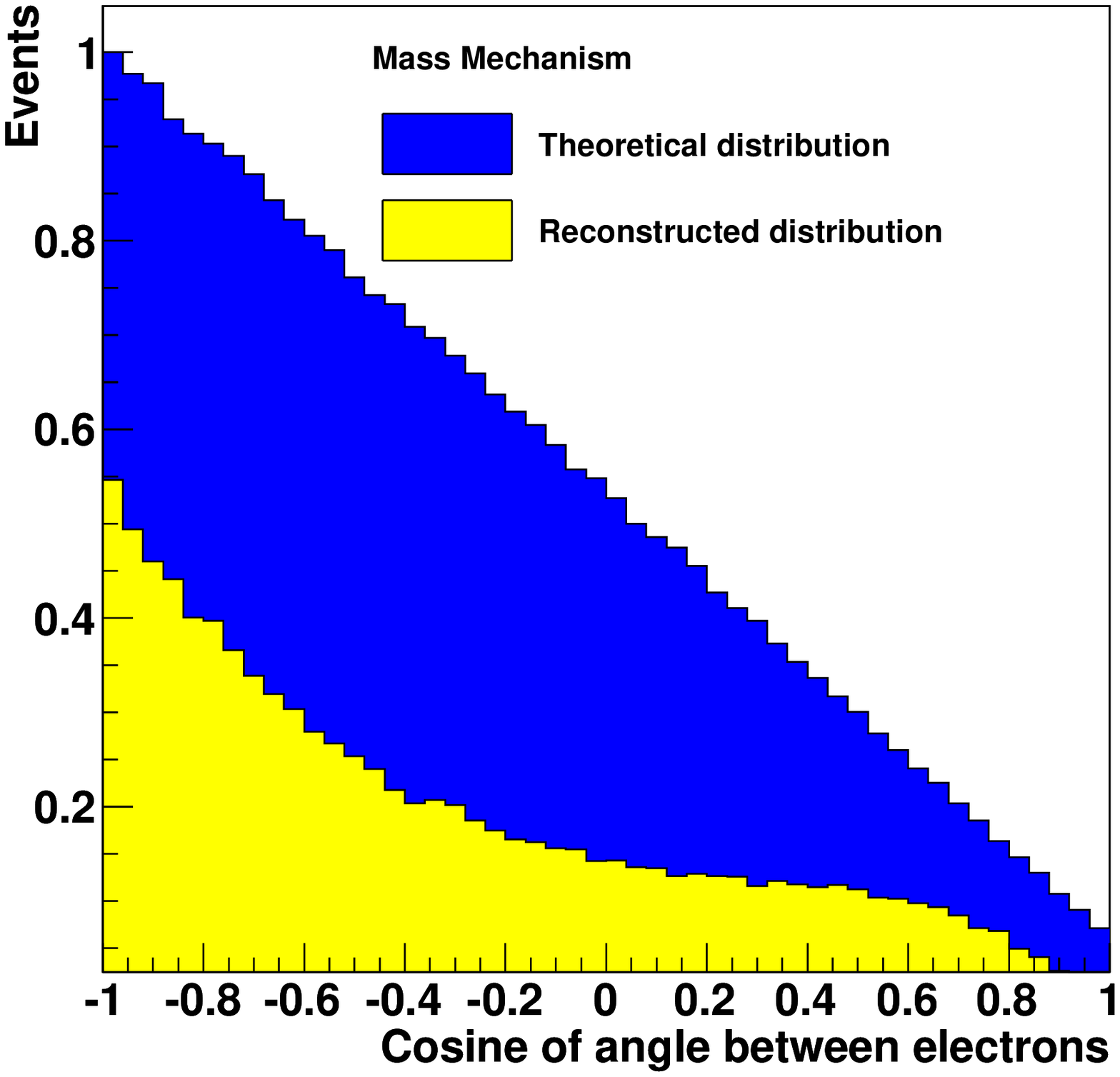}
\label{fig:dists-a}
}
\subfloat[][]{
\includegraphics[clip,width=0.37\textwidth]{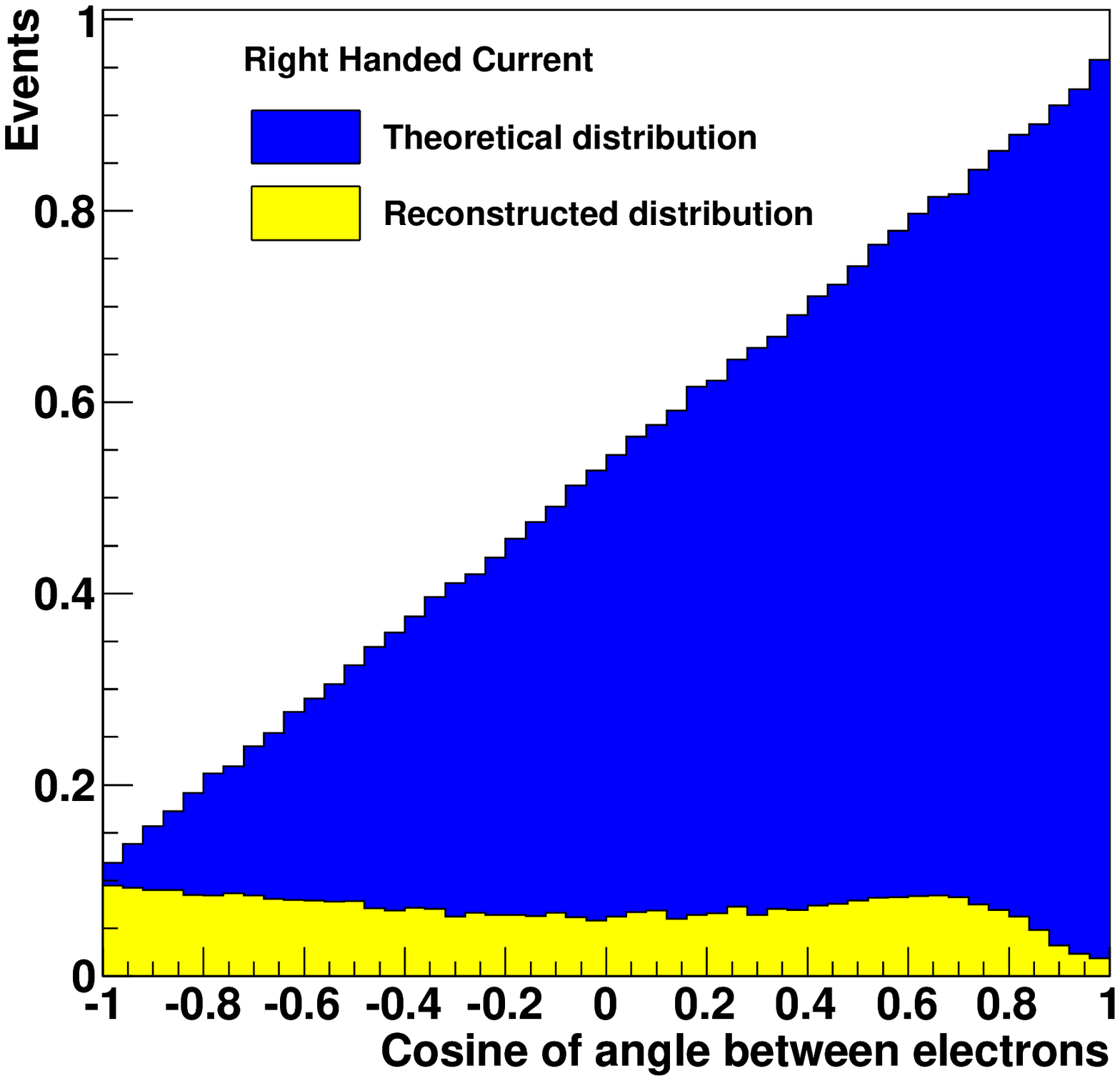}
\label{fig:dists-b}
}

\subfloat[][]{
\includegraphics[clip,width=0.37\textwidth]{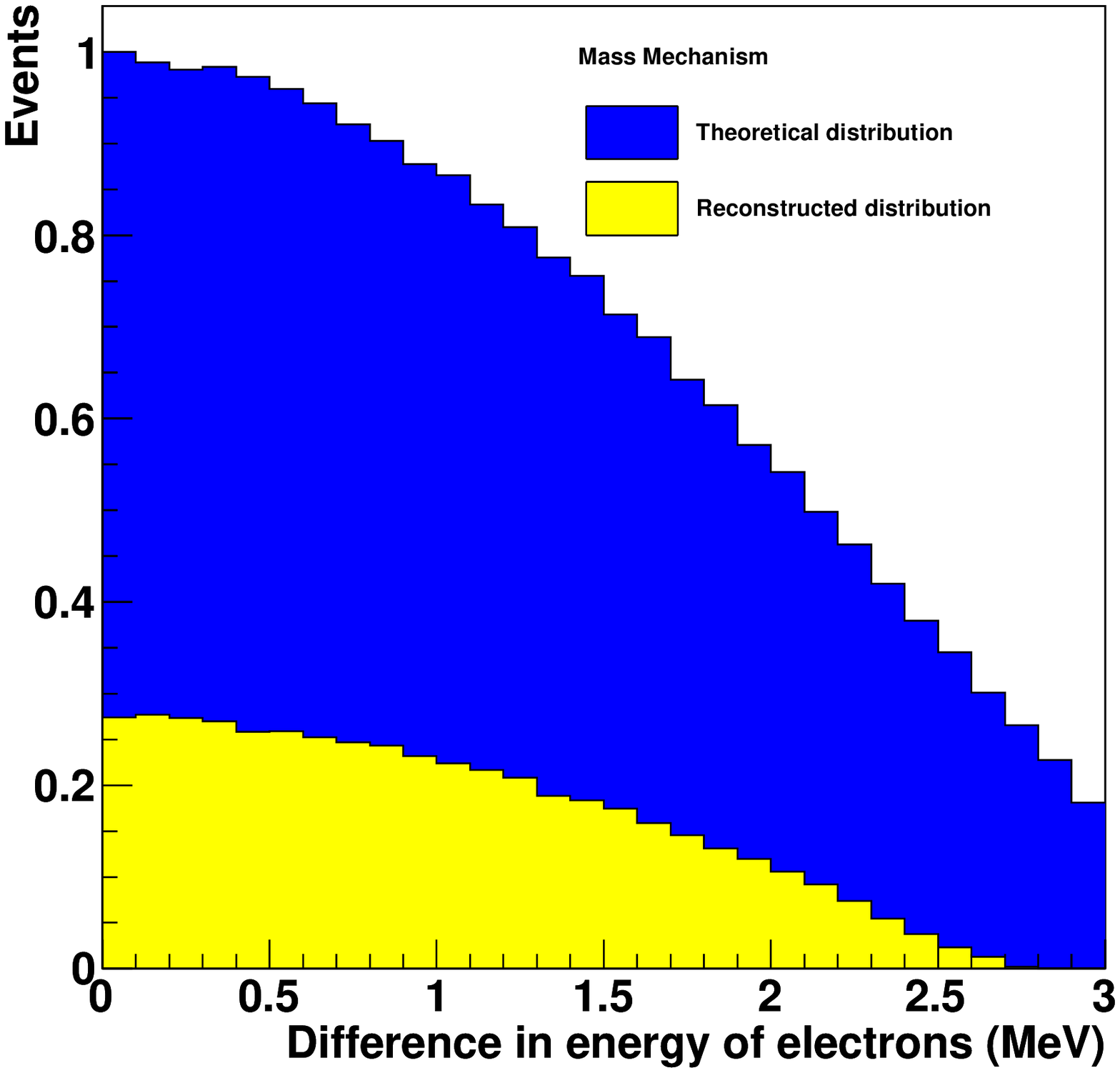}
\label{fig:dists-c}
}
\subfloat[][]{
\includegraphics[clip,width=0.37\textwidth]{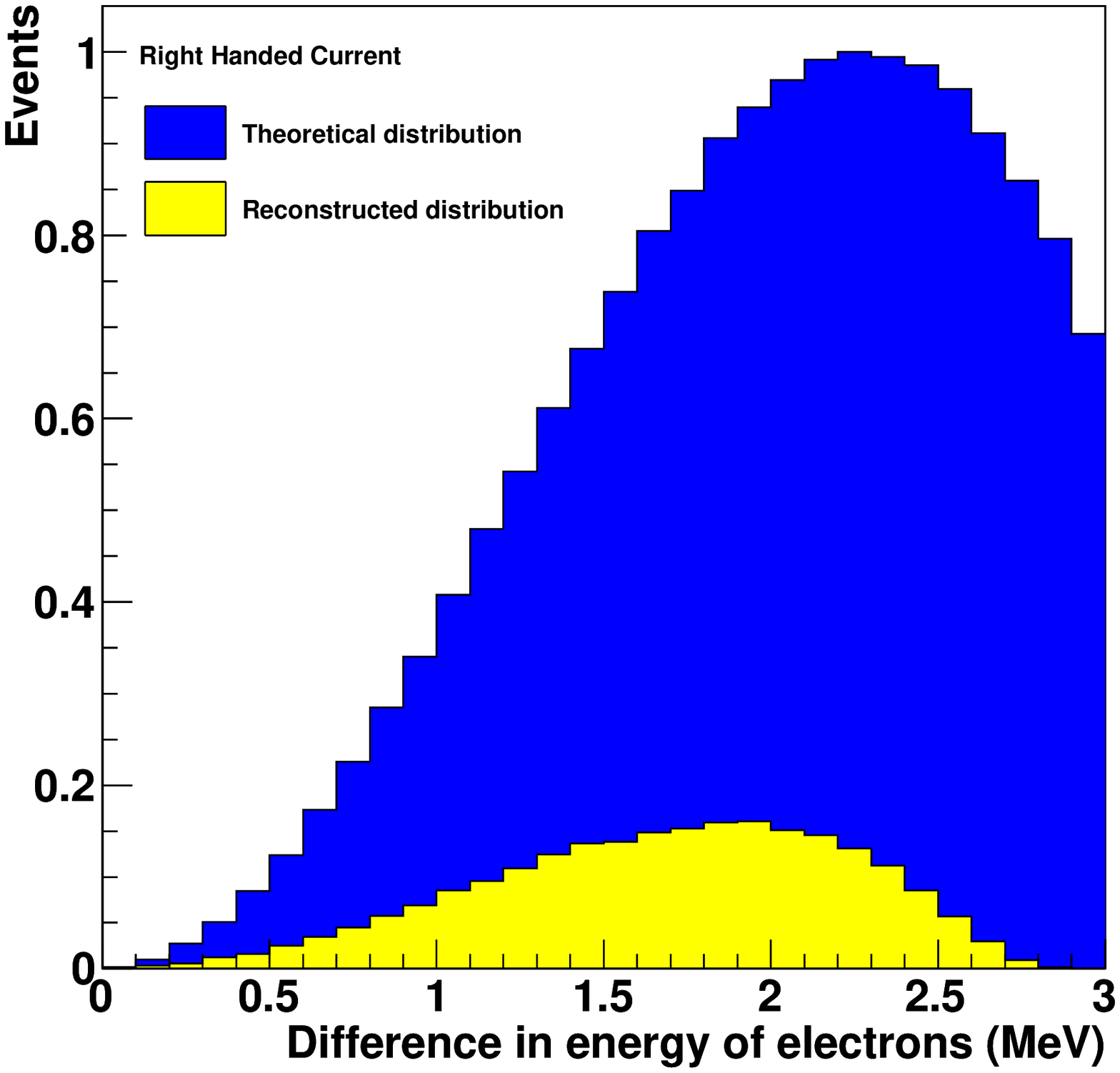}
\label{fig:dists-d}
}
\caption{Theoretical and experimental electron angular distributions for (a) MM and (b) RHC$_\lambda$. Theoretical and experimental electron energy difference distributions for (c) MM and (d) RHC$_\lambda$. All distributions are shown for the isotope \(^{82}\)Se and the reconstructed distributions are normalised to the theoretical distribution to show signal efficiency.} 
\label{fig:dists}
\end{figure*}

The variables reconstructed from the simulation are the energy sum, where a peak is expected at the energy release, $Q$, of the $0\nu\beta\beta$ decay, the energy difference and the cosine of the opening angle of the two electrons. Simulations of the angular and energy difference distributions of the two electrons in a signal sample are shown in Fig.~\ref{fig:dists} for the isotope $^{82}$Se (similar results hold for $^{150}$Nd). 
The reconstructed distributions, normalised to the theoretical distributions, show detector effects which arise due to multiple scattering in the source foil, compared to the theoretically predicted distributions based on Equations~(\ref{eq:DistroAngle}) and (\ref{eq:DistroEnergyDiff}). This influence is particularly strong in the right-handed current as one electron usually has low energy so the shape of the distribution is changed (on average a 30$^\circ$ deviation from the generated distribution). The reconstruction efficiency is also low for small angular separation between the electrons when they travel through the same drift cells. 

\begin{figure*}[!t]
\centering
\subfloat[][]{
\includegraphics[clip,width=0.45\textwidth]{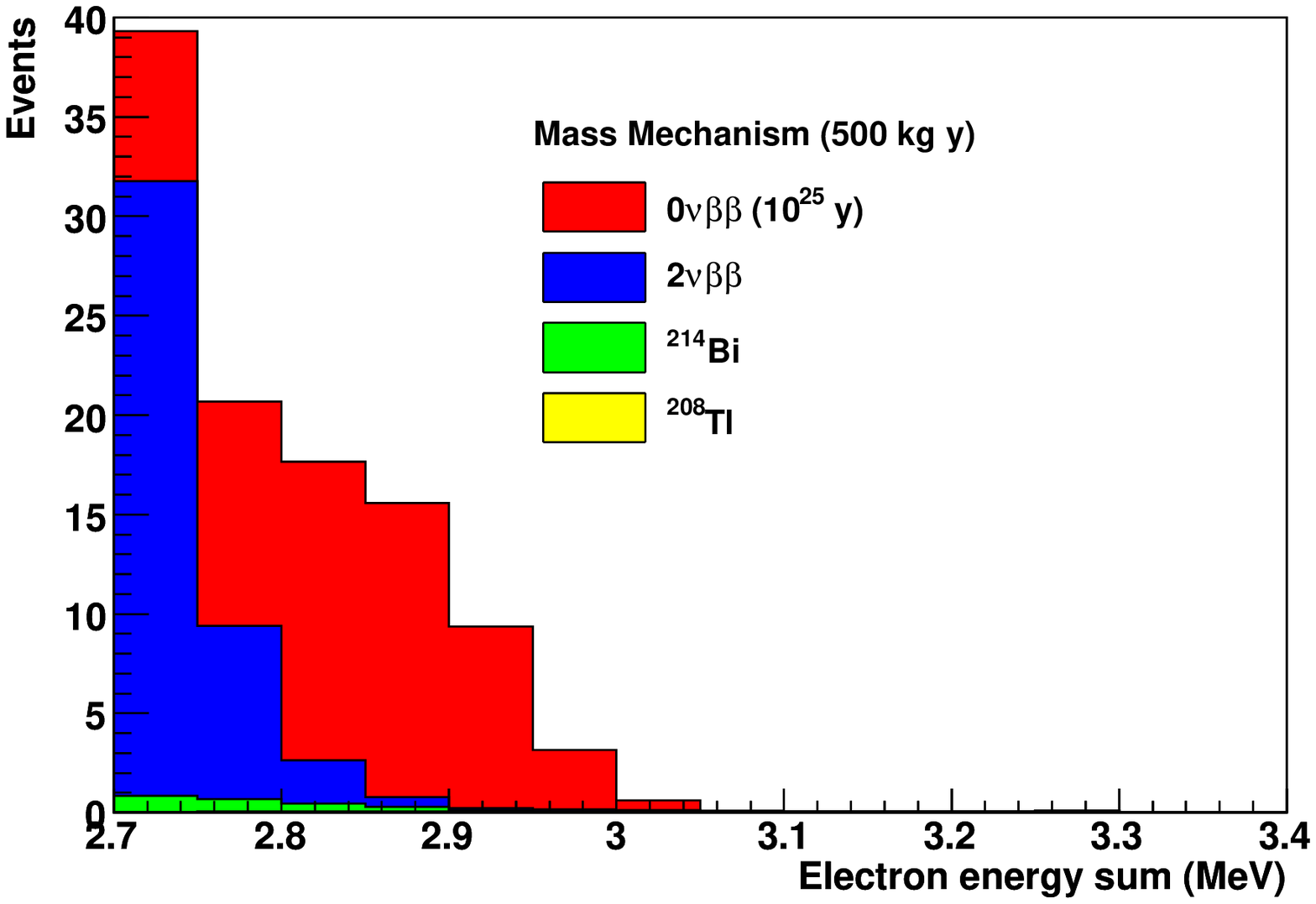}
}
\subfloat[][]{
\includegraphics[clip,width=0.45\textwidth]{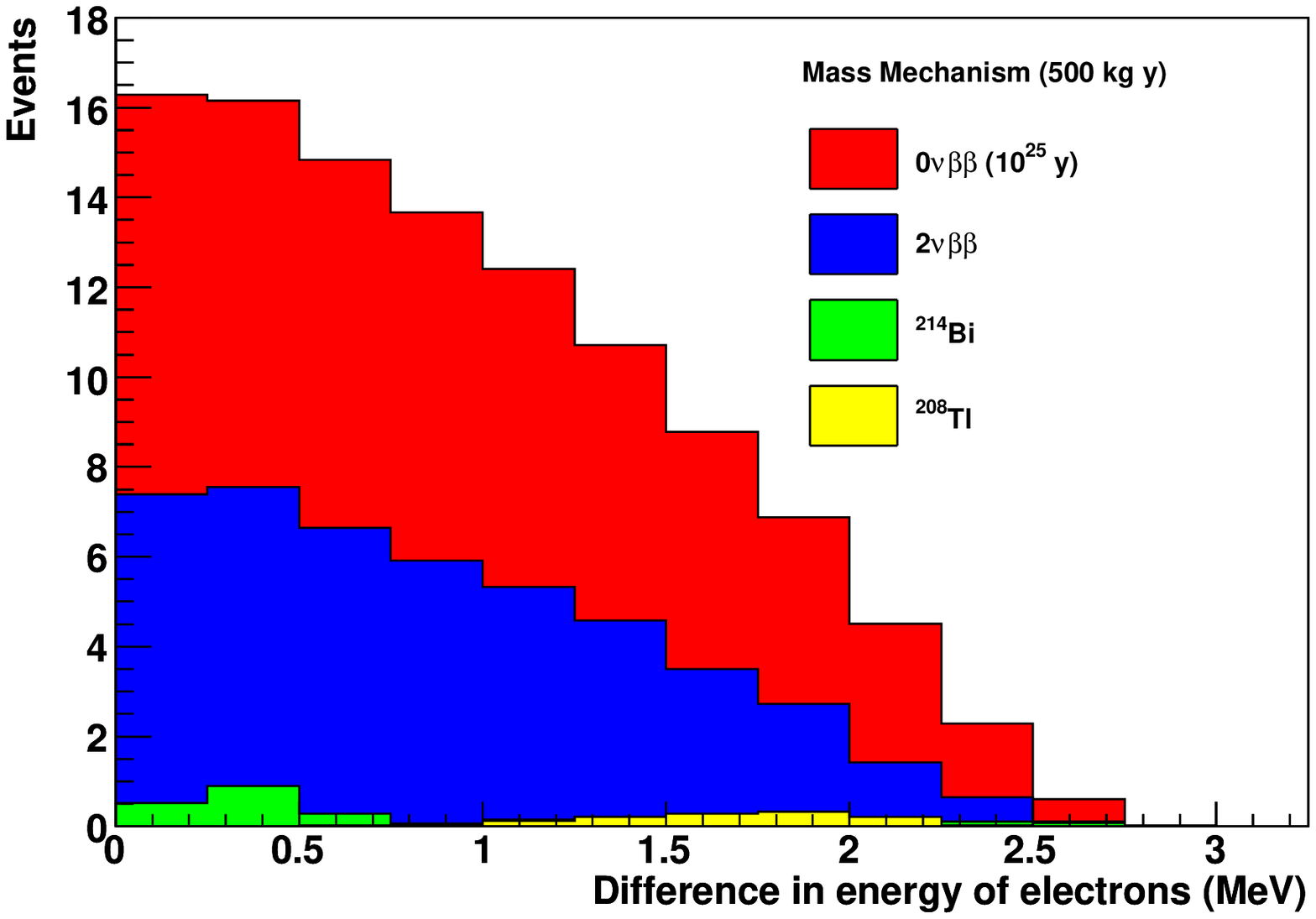}
}

\subfloat[][]{
\includegraphics[clip,width=0.45\textwidth]{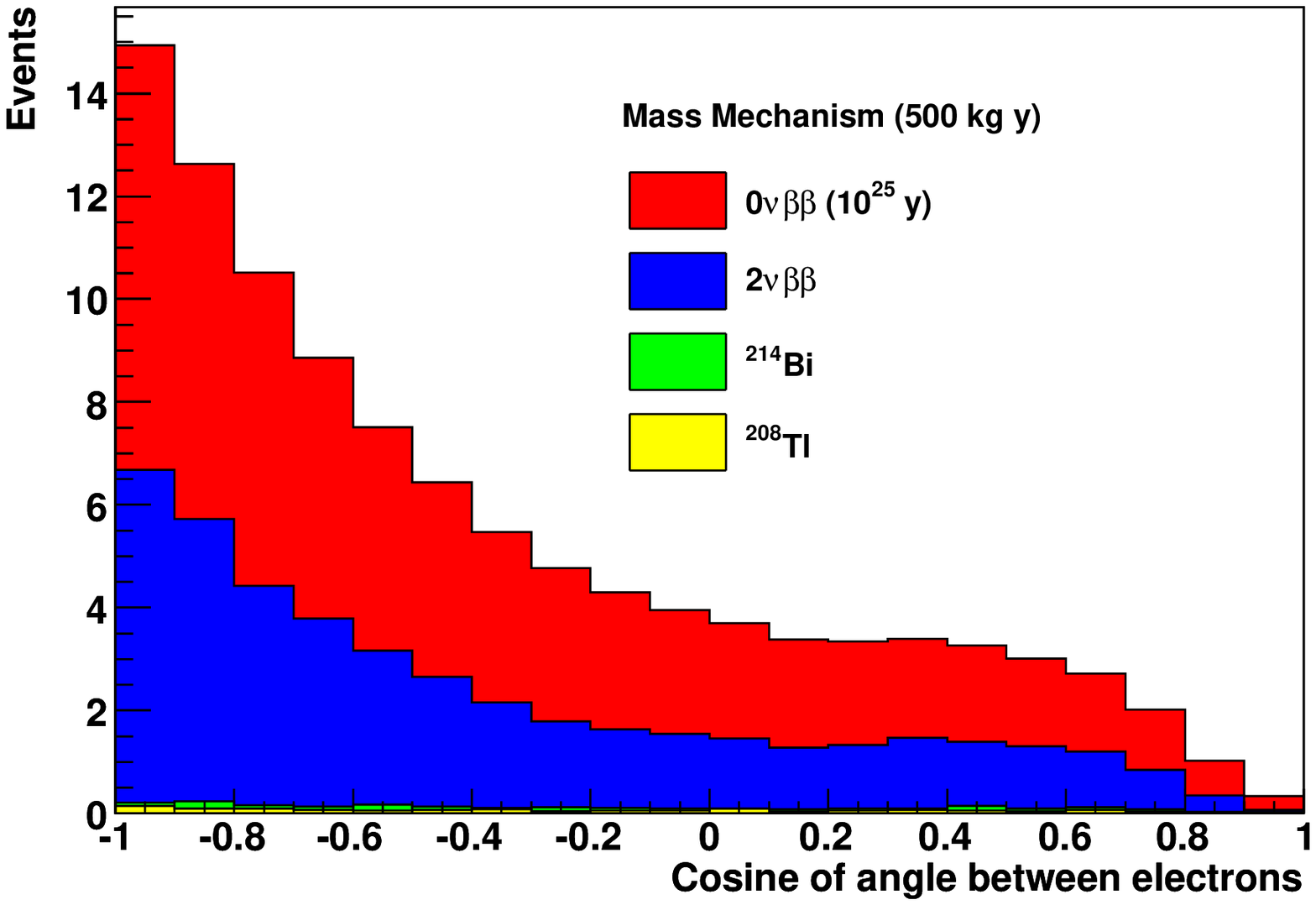}
}
\caption{Expected number of MM signal (half-life of $10^{25}$~y) and background events in $^{82}$Se after $500$~kg~y exposure shown for (a) electron energy sum, (b) electron energy difference and (c) cosine of angle between electrons.} 
\label{fig:dists_back}
\end{figure*}

The backgrounds were processed by the same detector simulation and reconstruction programs as the signal. The dominant two neutrino double $\beta$ decay ($2\nu\beta\beta$) background and the background due to foil contamination were normalised assuming a detector exposure of $500$~kg y. Due to the high decay energy $Q$ for $0\nu\beta\beta$ in $^{150}$Nd, the $^{214}$Bi background may be neglected. The activities were assumed to be $2$~$\mu$Bq/kg for $^{208}$Tl and $10$~$\mu$Bq/kg for $^{214}$Bi. These are the target radioactive background levels in the baseline SuperNEMO design. Reconstructed distributions of the experimental variables including background events for the MM at an example signal half-life of $10^{25}$~y are shown in Figure~\ref{fig:dists_back}.

\subsection{Limit Setting}\label{sec:ls}
\begin{figure*}[!t]
\centering
\includegraphics[clip,width=0.45\textwidth]{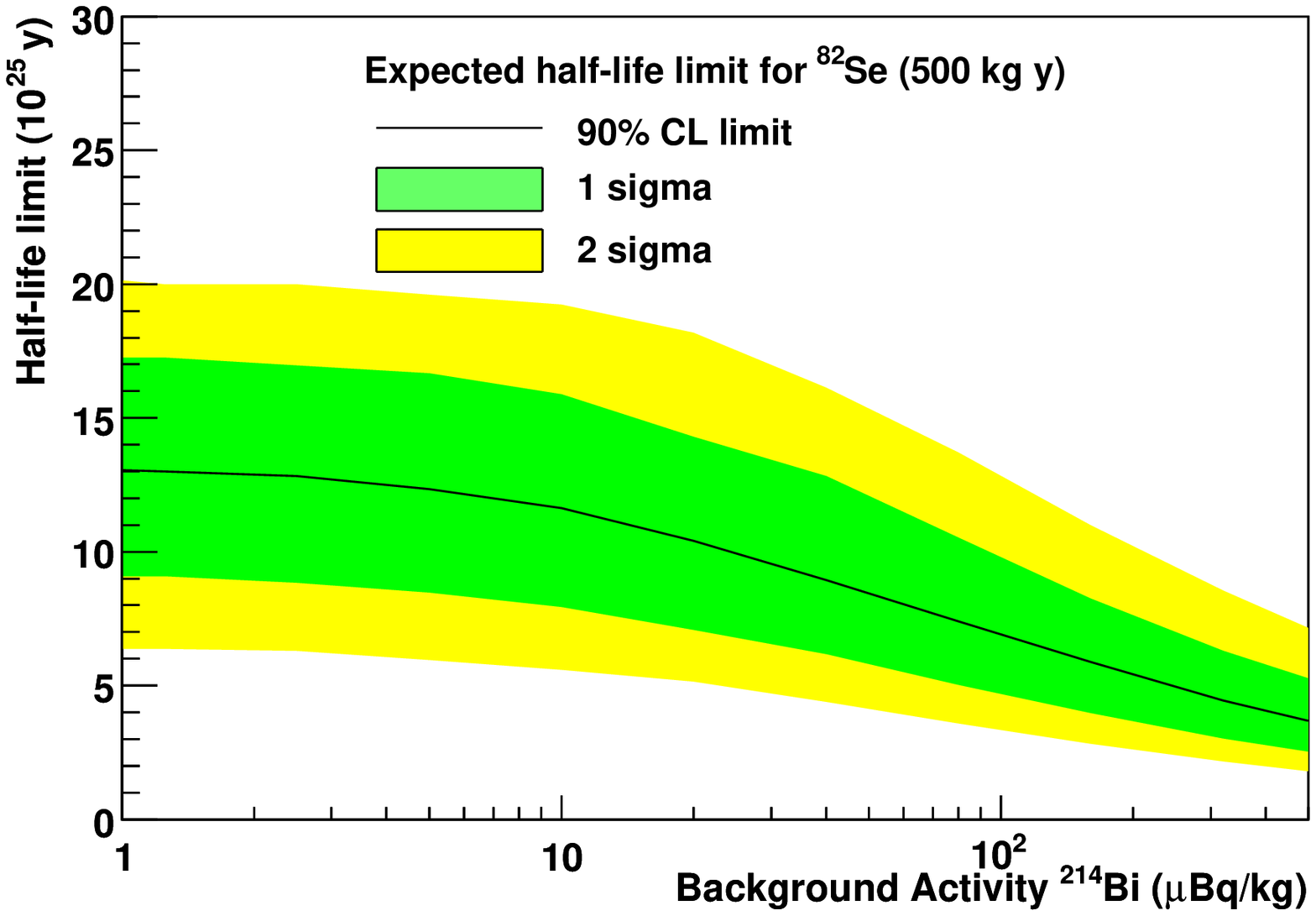}
\caption{Expected limit on the $0\nu\beta\beta$ half-life due to the MM for SuperNEMO under the background-only hypothesis. The expected limit with the one and two standard deviation bands is shown as a function of background activity for $^{214}$Bi in $^{82}$Se (a $^{208}$Tl activity of $2$~$\mu$Bq/kg is assumed).} 
\label{fig:limits_bkg}
\end{figure*}

To determine the longest half-life that can be probed with SuperNEMO, exclusion limits at $90\%$ CL on the half-life using the distribution of the sum of electron energies (Fig.~\ref{fig:dists_back} (a)) were set using a Modified Frequentist ($CL_{s}$)~\cite{cls} method. This method uses a log-likelihood ratio (LLR) of the signal-plus-background hypothesis and the background-only hypothesis, where the signal is due to the $0\nu\beta\beta$ process. The effect of varying the $^{214}$Bi background activities on the expected limit to the MM is shown in Fig.~\ref{fig:limits_bkg}. The expected limit is given by the median of the distribution of the LLR and the widths of the bands shown represent one and two standard deviations of the LLR distributions for a given $^{214}$Bi activity. For comparison, the NEMO-III internal $^{214}$Bi background is $<100$~$\mu$Bq/kg in $^{100}$Mo and $530 \pm 180$~$\mu$Bq/kg in $^{82}$Se. The NEMO-III internal $^{208}$Tl background is $110 \pm 10$~$\mu$Bq/kg in $^{100}$Mo, $340 \pm 50$~$\mu$Bq/kg in $^{82}$Se and $9320 \pm 320$~$\mu$Bq/kg in $^{150}$Nd~\cite{NEMO_background}. The $\gamma$-veto used reduces the number of radioactive background events by $30\%$ for $^{214}$Bi in the electron energy sum window $>2.7$~MeV. 

All external backgrounds from outside the foil, apart from radon in the tracking volume, are expected to be negligible and were not simulated. The energy distribution of the external radon background is similar to the internal background. Simulations have shown that a contamination of $10$~$\mu$Bq/kg of $^{214}$Bi in the foil is equivalent to $280$~$\mu$Bq/m$^{3}$ of $^{214}$Bi in the gas volume and $2$~$\mu$Bq/kg of $^{208}$Tl in the foil is equivalent to $26$~$\mu$Bq/m$^{3}$ of $^{208}$Tl in the gas volume. Figure~\ref{fig:limits_bkg} shows that this level of external background would lead to a $\sim15\%$ reduction in the half-life limit. The dominant $2\nu\beta\beta$ background is measured by SuperNEMO and statistical uncertainties on its half-life are expected to be negligible. Inclusion of an estimated $7\%$ correlated systematic uncertainty on the signal and background distributions~\cite{NEMO_Se82} leads to a $\sim5\%$ reduction in the MM half-life limit. The effects of external background and of systematic uncertainties on the $2\nu\beta\beta$ background are not included in the results of this paper.      

\begin{figure*}[!t]
\centering
\subfloat[][]{
\includegraphics[clip,width=0.43\textwidth]{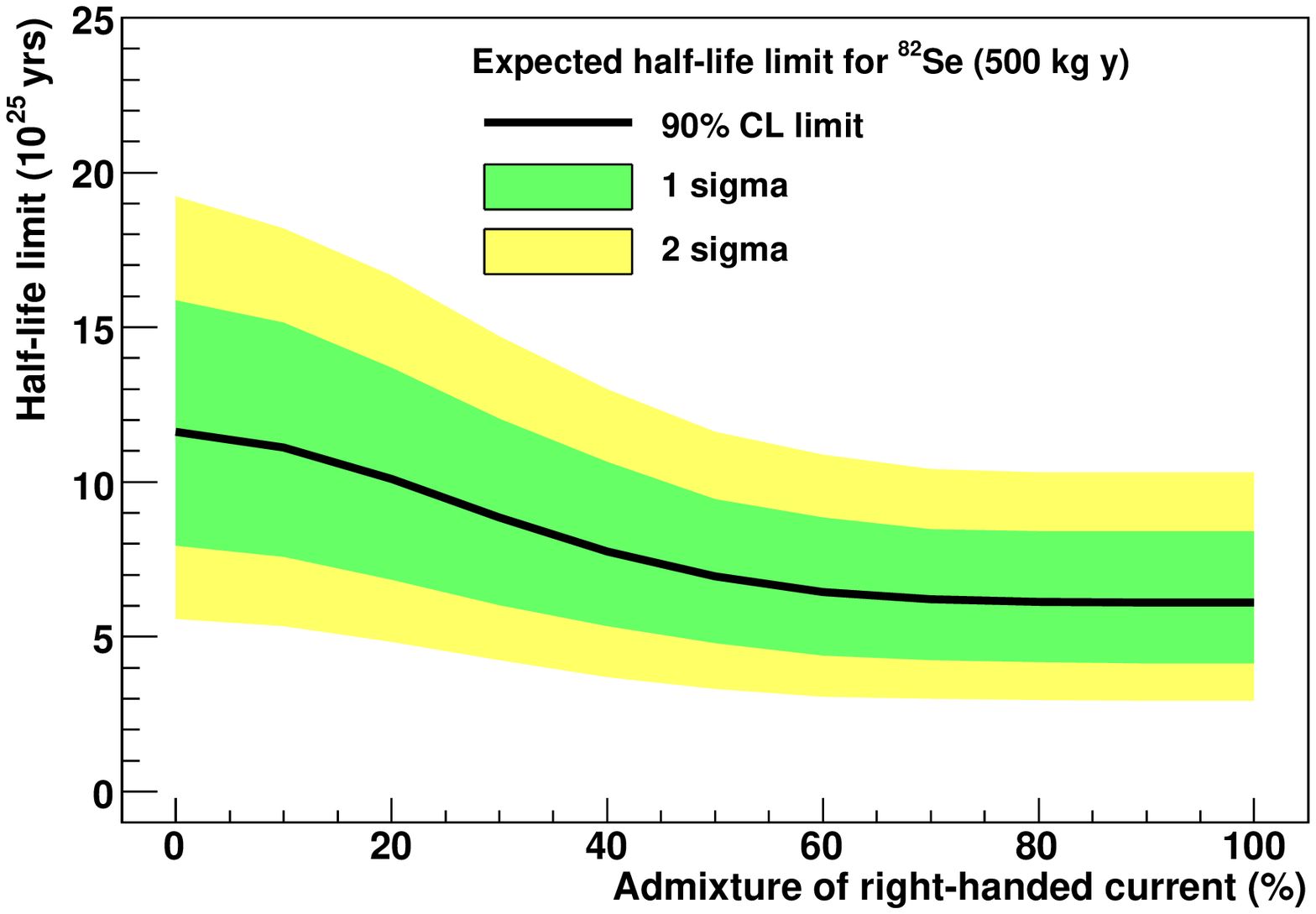}
\label{fig:limits_1}
}
\qquad
\subfloat[][]{
\includegraphics[clip,width=0.43\textwidth]{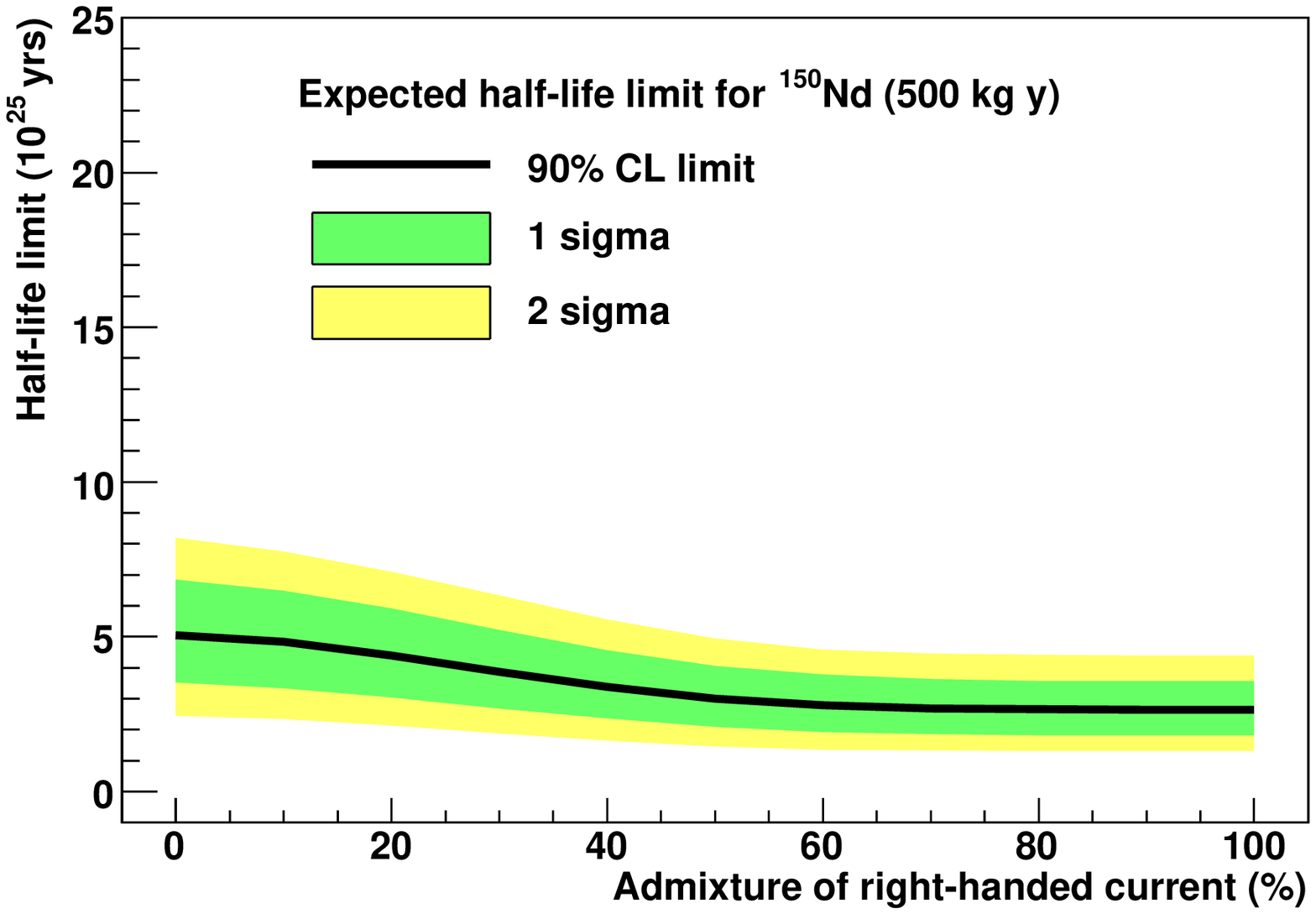}
\label{fig:limits_2}
}
\caption{Expected limit on the $0\nu\beta\beta$ half-life for SuperNEMO under the background-only hypothesis. The expected limit with the one and two standard deviation bands is shown as a function of admixture of the RHC$_\lambda$ mechanism for (a) $^{82}$Se and (b) $^{150}$Nd.} 
\label{fig:limits}
\end{figure*}

Expected exclusion limits at 90\% confidence level on the half-life are shown in Fig.~\ref{fig:limits}. Results are displayed as a function of RHC$_\lambda$ admixture, where the signal distribution is produced by combining weighted combinations per bin of the MM and RHC$_\lambda$ contributions at the event level. An admixture of $0\%$ corresponds to a pure MM contribution, and an admixture of $100\%$ to pure RHC$_\lambda$. Interference terms are assumed to be small and are neglected in the experimental simulation. The lower efficiency in the case of RHC$_\lambda$ results in a lower limit for larger admixtures. The half-life limit is approximately twice as sensitive in measurements of $^{82}$Se due to the lower mass number and higher $2\nu\beta\beta$~decay half-life, though this is compensated in $^{150}$Nd by more favourable phase space when converting into physics parameter space. In the case where one mechanism dominates SuperNEMO is expected to be able to exclude $0\nu\beta\beta$ half-lives up to $1.2\cdot10^{26}$~y (MM) and $6.1\cdot 10^{25}$~y (RHC$_\lambda$) for $^{82}$Se, and $5.1\cdot 10^{25}$~y (MM) and $2.6\cdot10^{25}$~y (RHC$_\lambda$) for $^{150}$Nd.

\subsection{Observation}\label{sec:obs}
\begin{figure*}[!t]
\centering
\subfloat[][]{
\includegraphics[clip,width=0.43\textwidth]{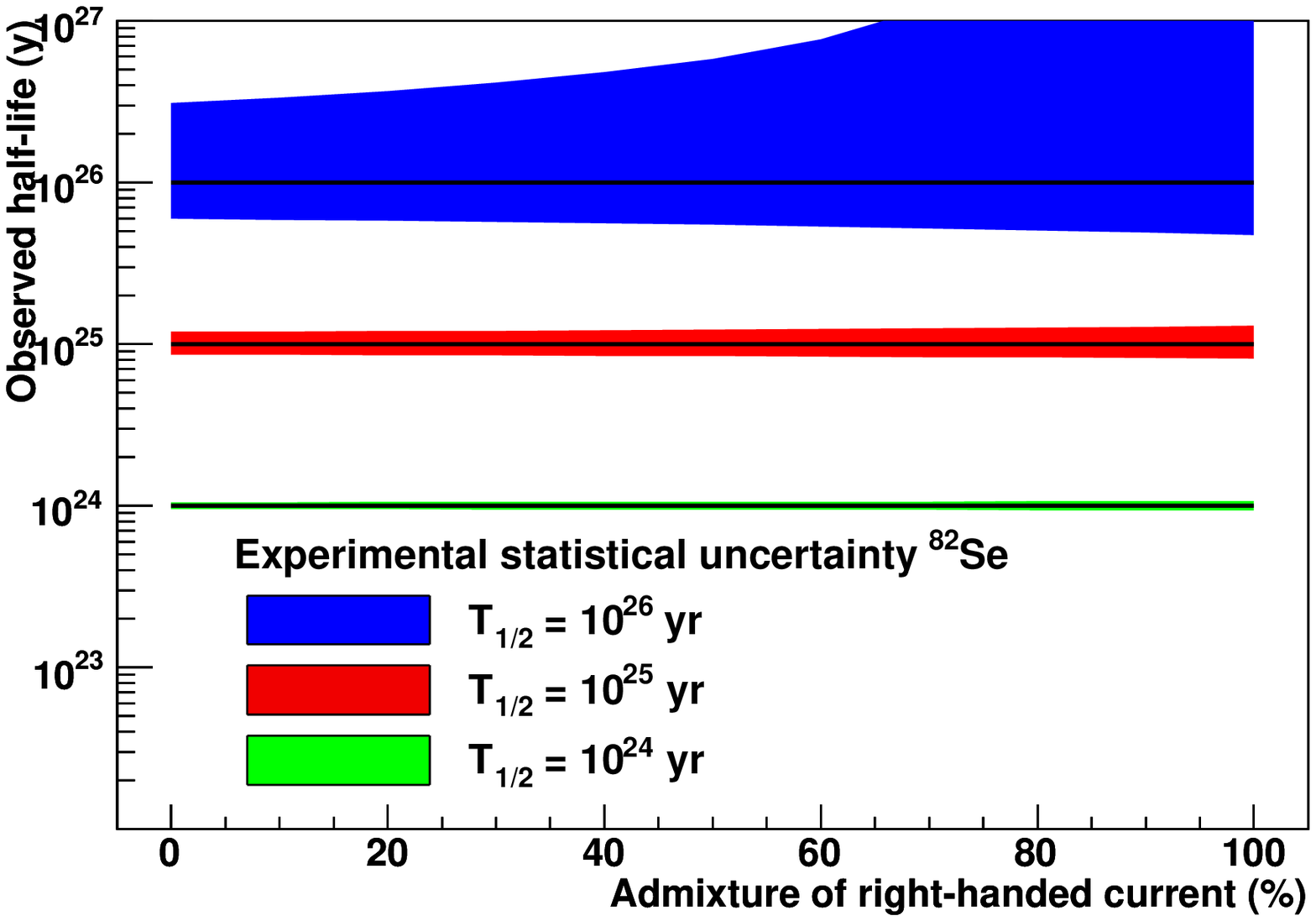}
\label{fig:staterr_1}
}
\qquad
\subfloat[][]{
\includegraphics[clip,width=0.43\textwidth]{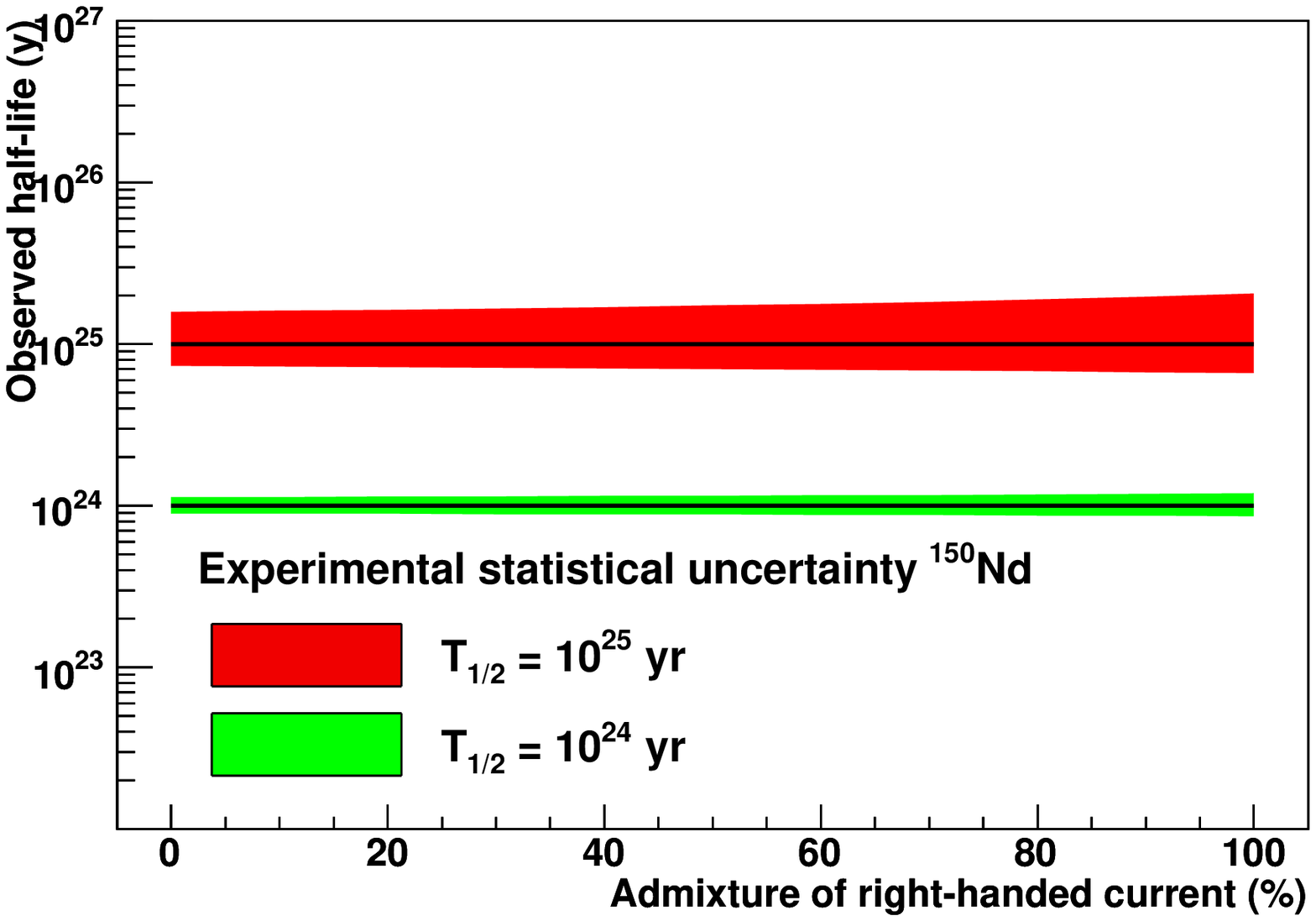}
\label{fig:staterr_2}
}
\caption{One standard deviation statistical uncertainties in the measurement of double $\beta$ decay half-lives at SuperNEMO as a function of admixture of the RHC$_\lambda$ mechanism represented as band thickness for (a) $^{82}$Se and (b) $^{150}$Nd.} 
\label{fig:staterr}
\end{figure*}

A $0\nu\beta\beta$ signal rate with significant excess over the background expectation, as for example shown in Fig.~\ref{fig:dists_back}, would lead to an observation. The expected experimental statistical uncertainties on the decay half-life are calculated from the Gaussian uncertainties on the observed number of signal and background events in the simulation. Figure~\ref{fig:staterr} shows the results for $^{82}$Se and $^{150}$Nd as a function of the admixture of RHC$_\lambda$. Acceptance effects cause the uncertainty to increase with admixture of RHC$_\lambda$. The statistical uncertainty increases significantly for large admixtures of RHC$_\lambda$ at $T_{1/2}=10^{26}$~y which go beyond the exclusion limit of SuperNEMO.

The angular asymmetry parameter $k_{\theta}$ in Equation~(\ref{eq:k}) is experimentally accessible by defining N$_{+}$ as the number of events with measured angle $\cos\theta<0$ and N$_{-}$ as the number of events with $\cos\theta>0$. Similarly, an energy difference asymmetry $k_{E}$ can be obtained where N$_{+}$ is the number of events with energy difference $<$ $Q/2$ (half the energy of the $0\nu\beta\beta$ decay) and N$_{-}$ is the number of events with energy difference $>$ $Q/2$. The electron energy sum is required to be greater than 2.7 MeV for $^{82}$Se and 3.1 MeV for $^{150}$Nd to maximise signal to background ratio. This results in signal efficiencies of 23.2\% for the MM and 13.2\% for the RHC$_{\lambda}$ in $^{82}$Se and 19.1\% for the MM and 10.4\% for the RHC$_{\lambda}$ in $^{150}$Nd. 

\begin{figure*}[!t]
\centering
\subfloat[][]{
\includegraphics[clip,width=0.45\textwidth]{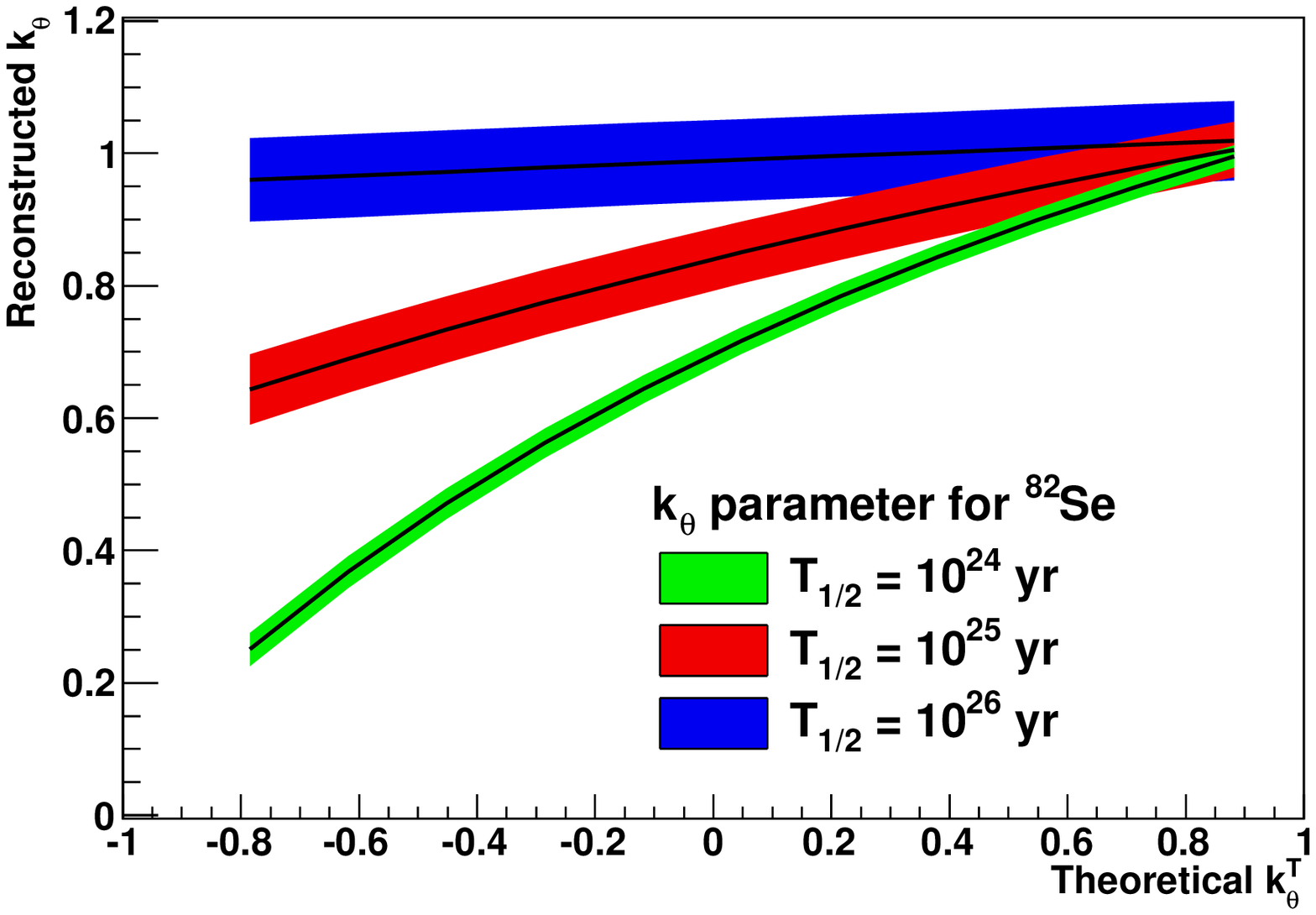}
}
\subfloat[][]{
\includegraphics[clip,width=0.45\textwidth]{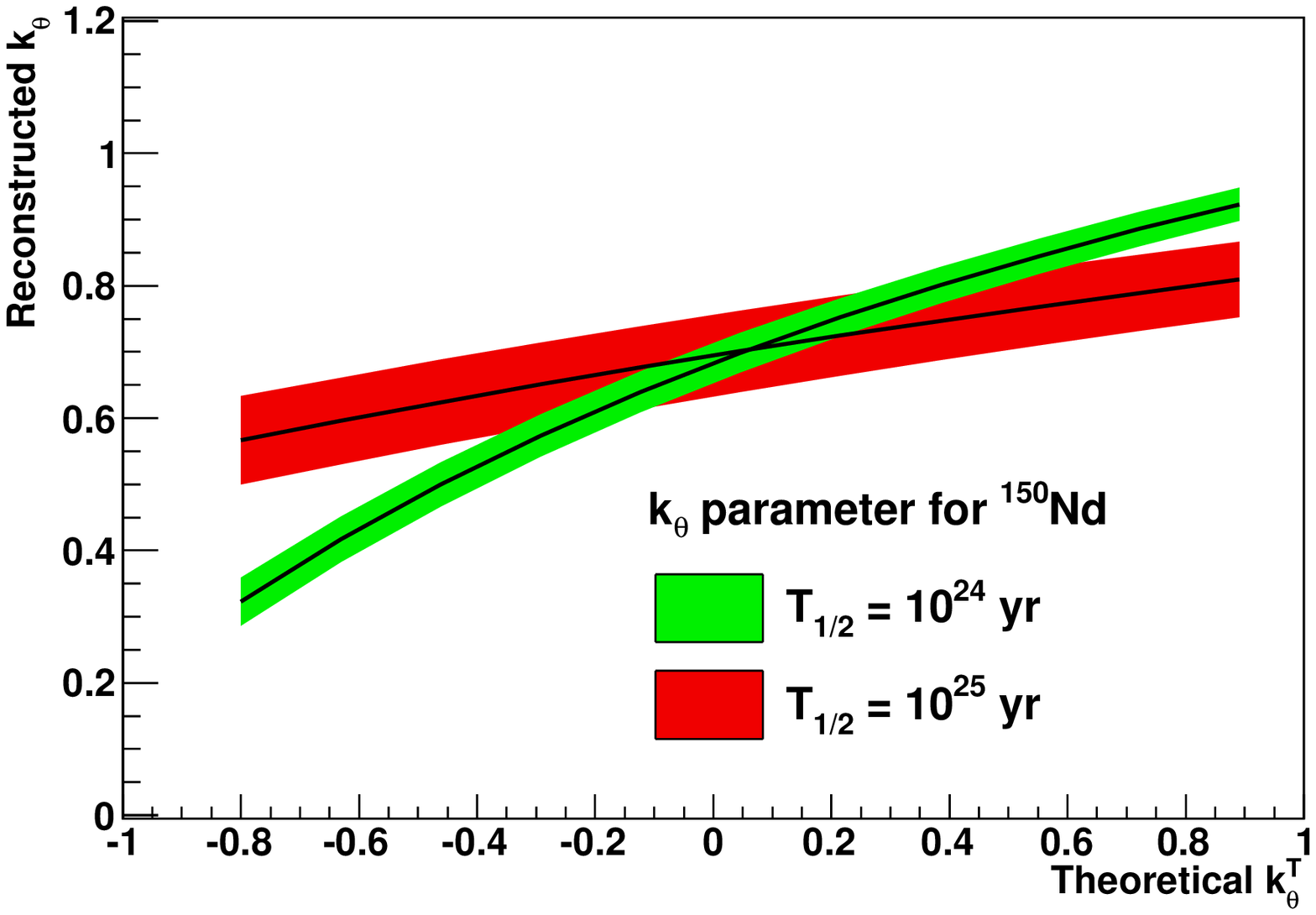}
}

\subfloat[][]{
\includegraphics[clip,width=0.45\textwidth]{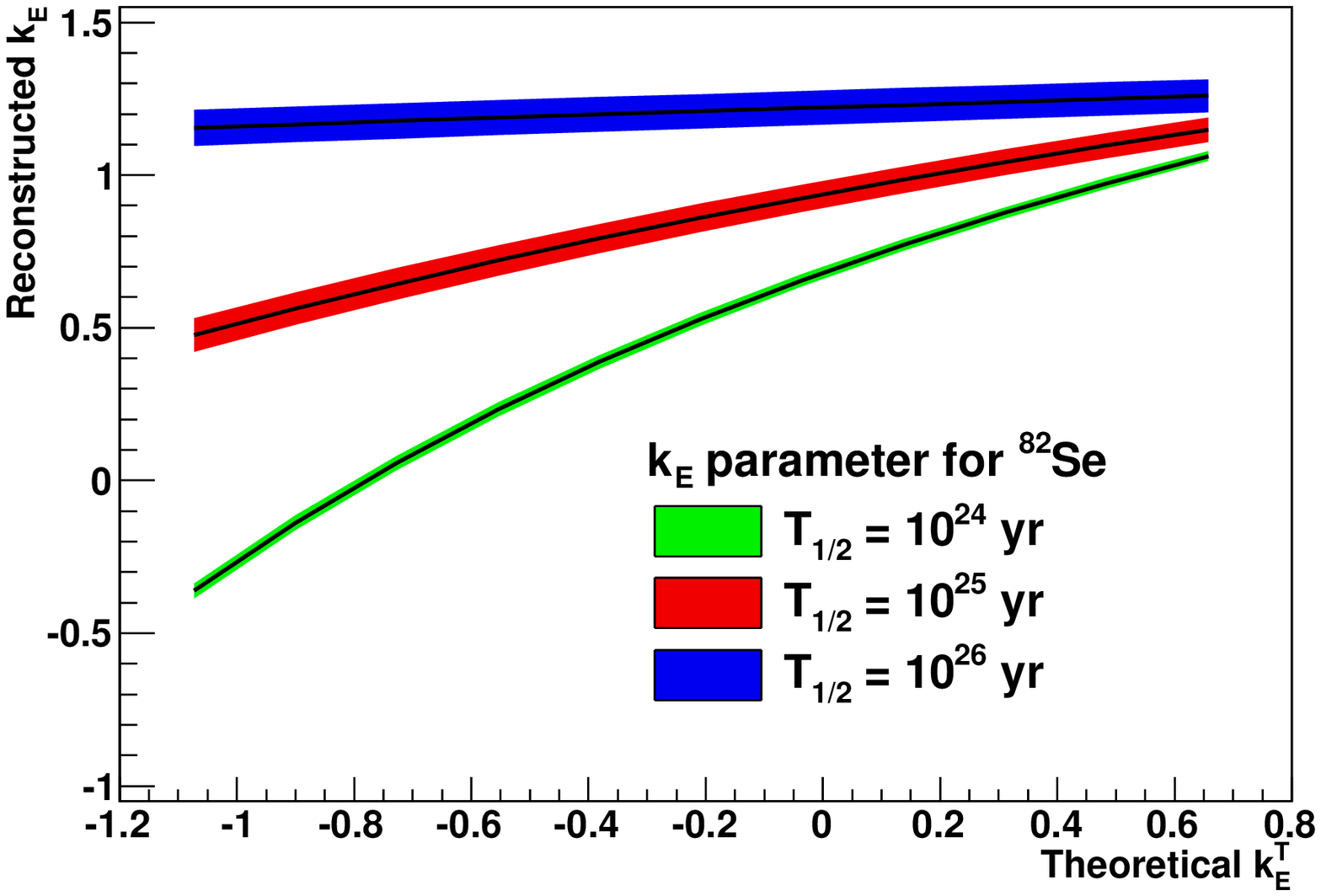}
}
\subfloat[][]{
\includegraphics[clip,width=0.45\textwidth]{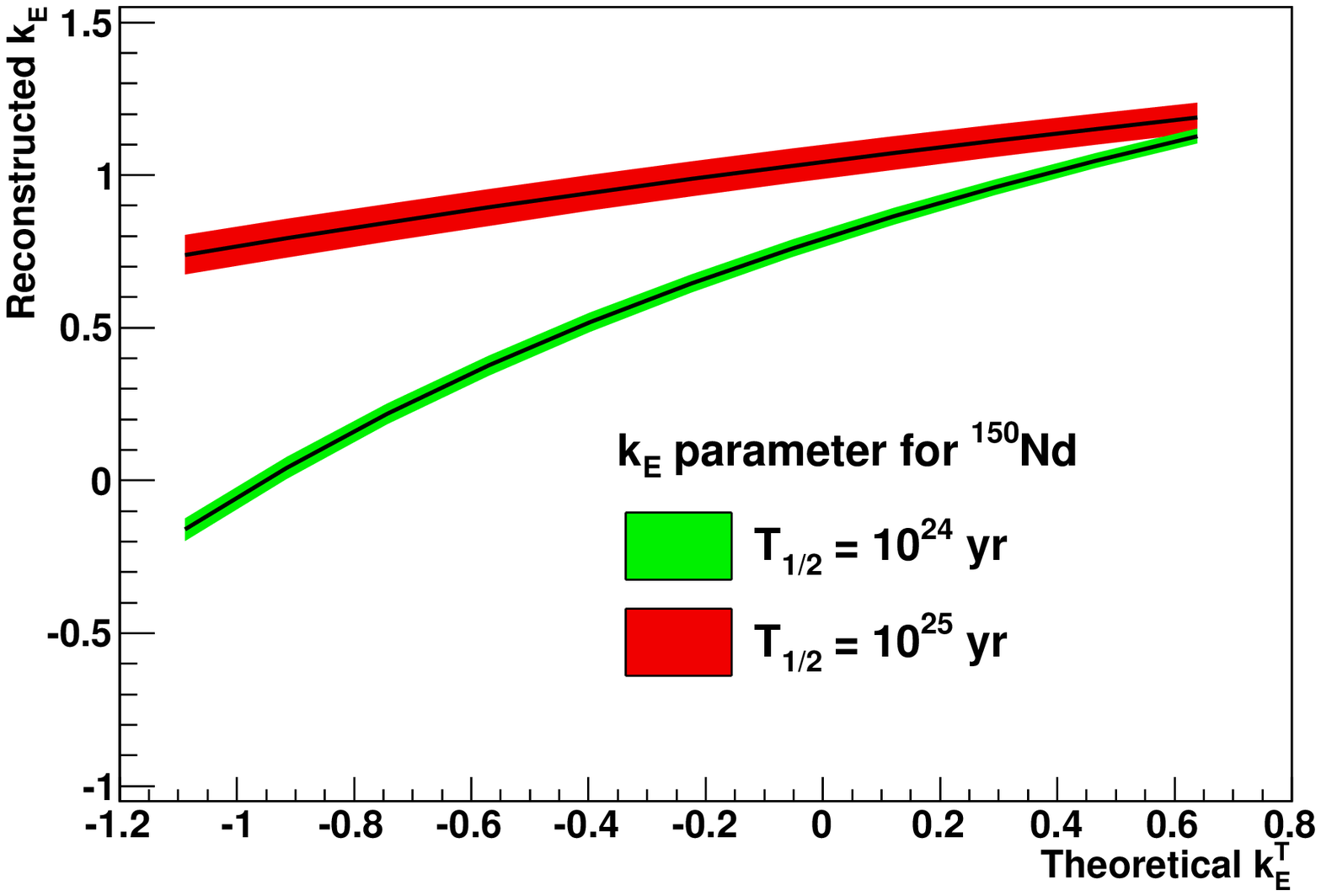}
}
\caption{Simulation of the correlation coefficients $k_\theta$ and $k_E$ as a function of theoretical $k_{\theta,E}^{T}$. The bands represent the one standard deviation statistical uncertainties. Shown are the angular correlation factor $k_\theta$ for $^{82}$Se (a) and $^{150}$Nd (b) and the energy difference correlation factor $k_E$ for $^{82}$Se (c) and $^{150}$Nd (d).} 
\label{fig:ktrue_krecon}
\end{figure*}

Experimentally, the distributions are only available as a sum of signal plus background so the measured values differ from the theoretically expected values due to the background distributions. This generally results in reconstructed correlation factors that are biased towards positive values. The measured values of $k_{\theta,E}$ are shown in Fig.~\ref{fig:ktrue_krecon} for a number of half-lives in the two isotopes. Statistical uncertainties are shown as the width of the bands. All reconstructed $k_{\theta,E}$ values are displayed as a function of the corresponding theoretical $k^{T}_{\theta,E}$ parameter, to allow for a model independent generalisation. It can be seen that the energy difference distribution allows stronger model discrimination than the angular distribution. 
%

%----------------------------------------------------------------------
\section{Probing New Physics}\label{sec:results}
\subsection{Model Parameter Constraints} \label{sec:RateSensitivity}

Having performed a detailed experimental analysis including a realistic simulation of the detector setup, the results are interpreted in terms of the expected reach of the SuperNEMO experiment to new physics parameters of the combined MM and RHC$_\lambda$ model of $0\nu\beta\beta$ decay.

\begin{figure*}[!t]
\centering
\subfloat[][]{
\includegraphics[clip,width=0.44\textwidth]{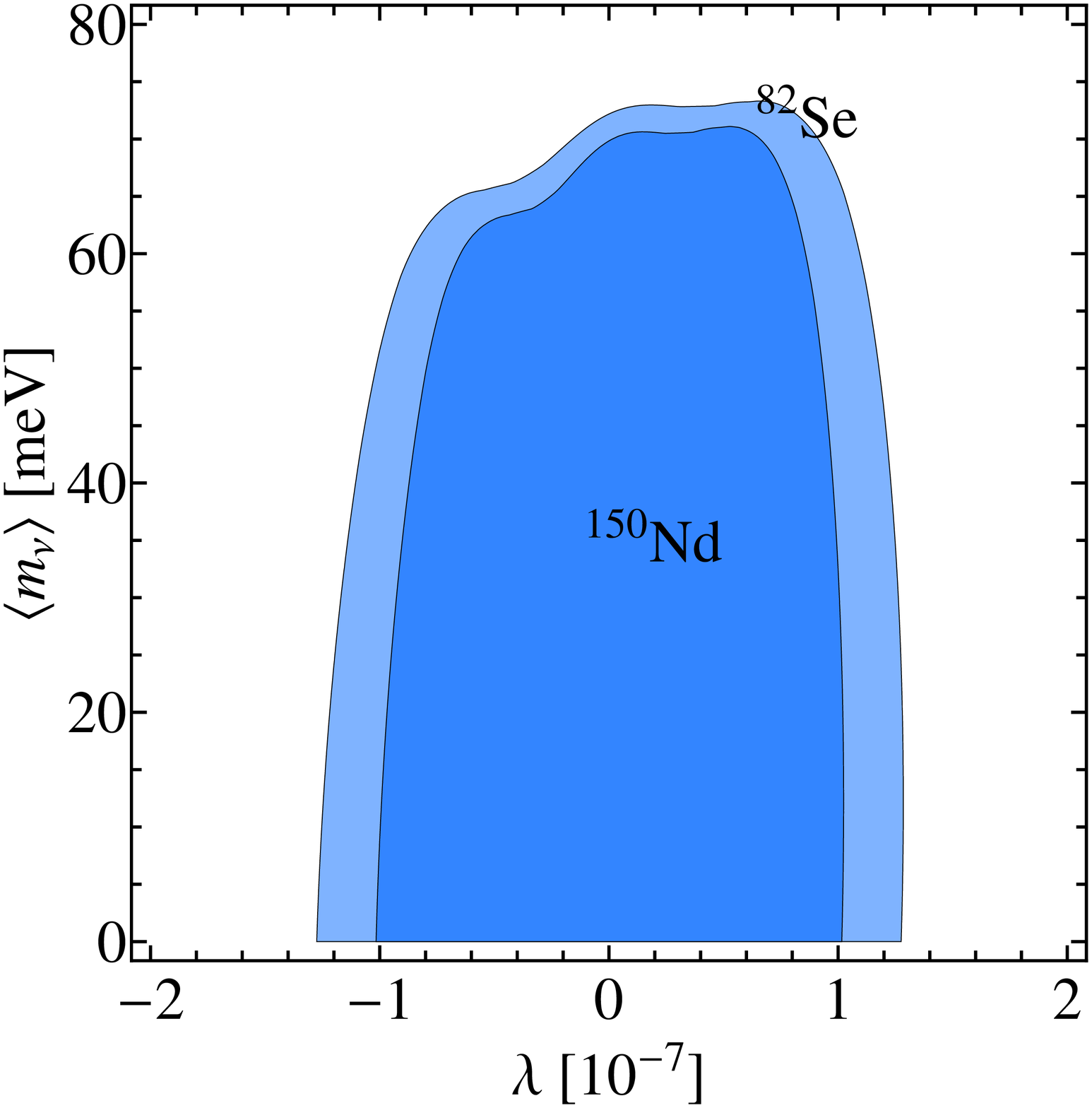}
}
\subfloat[][]{
\includegraphics[clip,width=0.46\textwidth]{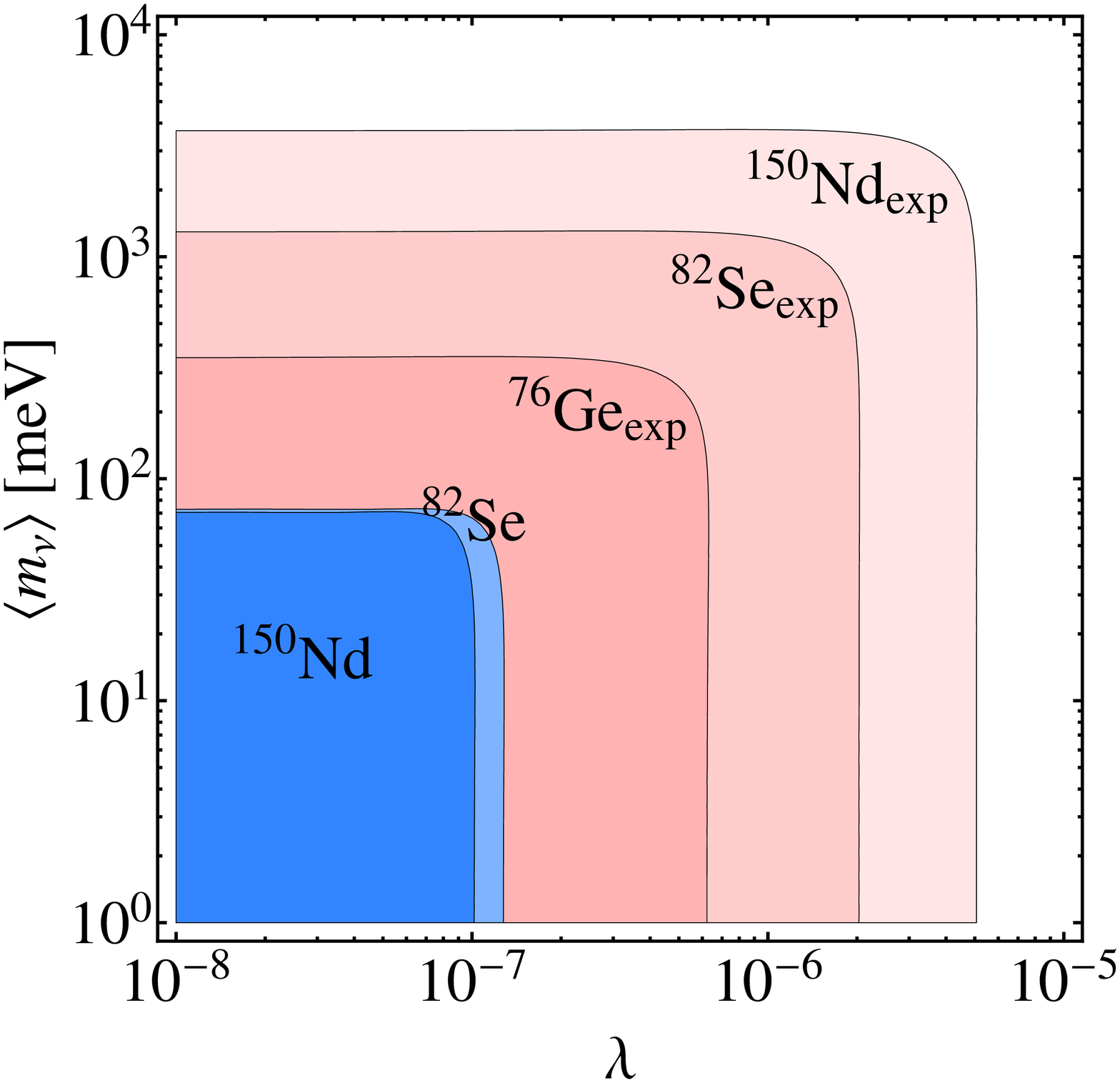}
}
\caption{(a) Expected SuperNEMO constraints on the model parameters $(m_\nu, \lambda)$ for the isotopes $^{82}$Se (light blue contour) and $^{150}$Nd (dark blue contour). (b) Comparison with current bounds on $0\nu\beta\beta$ half-lives of the isotopes $^{82}$Se (NEMO-III~\cite{NEMO_latestconf}), $^{150}$Nd (NEMO-III~\cite{NEMO_Nd150}) and $^{76}$Ge (Heidelberg Moscow~\cite{Heidelberg-Moscow}). The contours show the $90\%$ CL exclusion region.}
\label{fig:exclusion}
\end{figure*}
Using Equation~(\ref{eq:T12_mu_lambda}) for the $0\nu\beta\beta$ decay half-life together with the coefficients listed in Table~\ref{tab:NME}, the expected $90\%$ CL limit on $T_{1/2}$ shown in Fig.~\ref{fig:limits} can be translated into a constraint on the model parameters $m_\nu$ and $\lambda$. Assuming all other contributions are negligible this is shown in Fig.~\ref{fig:exclusion} (a), as a contour in the $m_\nu-\lambda$ parameter plane. In the case SuperNEMO does not see a signal these parameters would be constrained to be located within the coloured contour. The odd shape of the coloured contour is a direct consequence of the SuperNEMO $90\%$~CL exclusion limit as a function of the specific admixture between the MM and the RHC$_\lambda$ shown in Fig.~\ref{fig:limits}. The small interference term, though not included in the experimental simulation, is taken into account through Equation~(\ref{eq:T12_mu_lambda}) in this figure and results in the asymmetry of the distribution with respect to the sign of the parameter $\lambda$.  

As shown in Section~\ref{sec:simulation}, SuperNEMO is expected to be more sensitive to the $0\nu\beta\beta$ half-life when using the isotope $^{82}$Se, but this is compensated by the larger phase space of $^{150}$Nd. As a result, the constraint on the model parameters is slightly stronger for $^{150}$Nd. Due to the large uncertainty in the NMEs, this might be different for other NME calculations. To demonstrate the improvement over existing experimental bounds, the parameter constraints are shown in Fig.~\ref{fig:exclusion} (b) on a logarithmic scale (for positive values of $\lambda$), comparing the expected SuperNEMO reach with the current constraints from the $0\nu\beta\beta$ experiments NEMO-III~\cite{NEMO_latestconf,NEMO_Nd150} and Heidelberg Moscow \cite{Heidelberg-Moscow}.

Figure~\ref{fig:exclusion} shows that SuperNEMO is expected to constrain model parameters at $90\%$ CL down to $\langle m_\nu \rangle$=70-73 meV and $\lambda$=(1-1.3)$\cdot10^{-7}$. This would be an improvement by a factor 5-6 over the current best limit from the Heidelberg Moscow experiment and more than an order of magnitude compared to the NEMO-III results.  

%------------------------------------------------------------------------------
\subsection{Angular and Energy Correlations}\label{sec:angular}
As a more intriguing scenario it is now assumed that SuperNEMO actually observes a $0\nu\beta\beta$ decay signal in $^{82}$Se or $^{150}$Nd. Because of the tracking abilities described in Section~\ref{sec:simulation} this opens up the additional possibility of measuring the angular and energy distribution of the decays. Depending on the number of signal events detected, this can be crucial in distinguishing between different $0\nu\beta\beta$ decay mechanisms. In the analysis a reconstruction of the angular and energy correlation coefficients $k_{\theta}$ and $k_E$ is used to determine the theoretical coefficients, and thereby the admixture between the left- and right-handed currents in the combined MM and RHC$_\lambda$ model.

\begin{figure*}[!t]
\centering
\subfloat[][]{
\includegraphics[clip,width=0.30\textwidth]{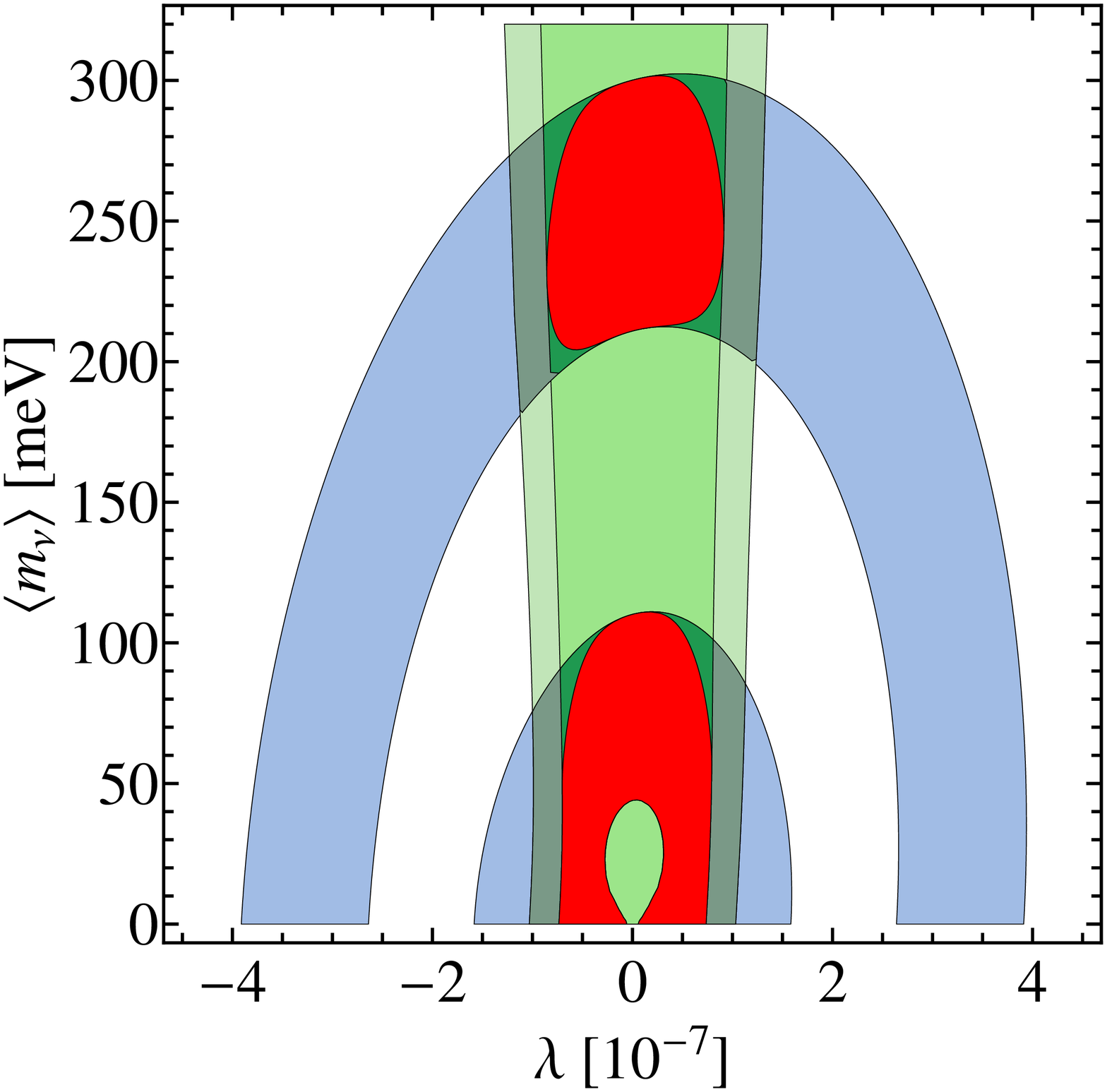}
}
\subfloat[][]{
\includegraphics[clip,width=0.30\textwidth]{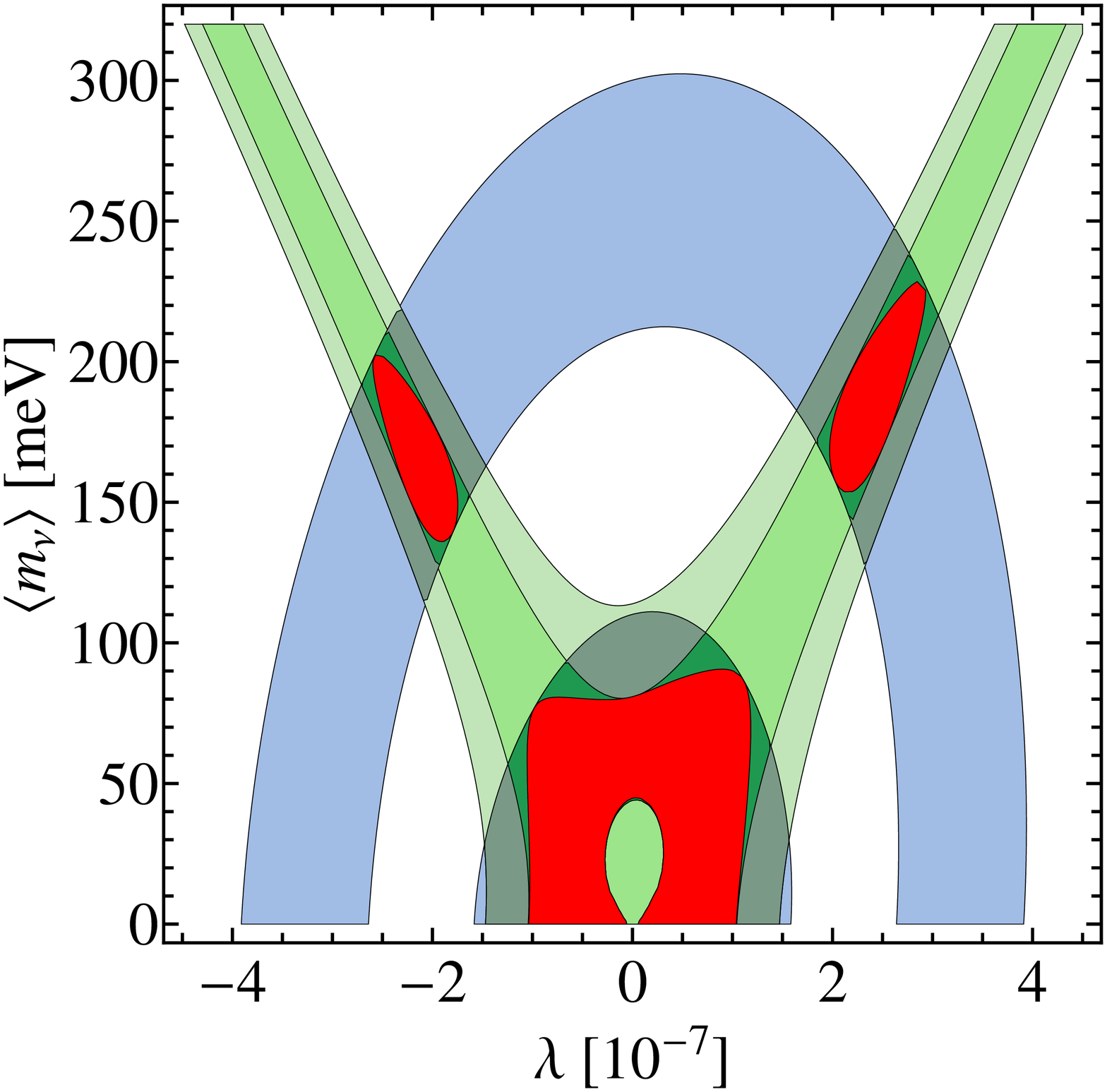}
}
\subfloat[][]{
\includegraphics[clip,width=0.30\textwidth]{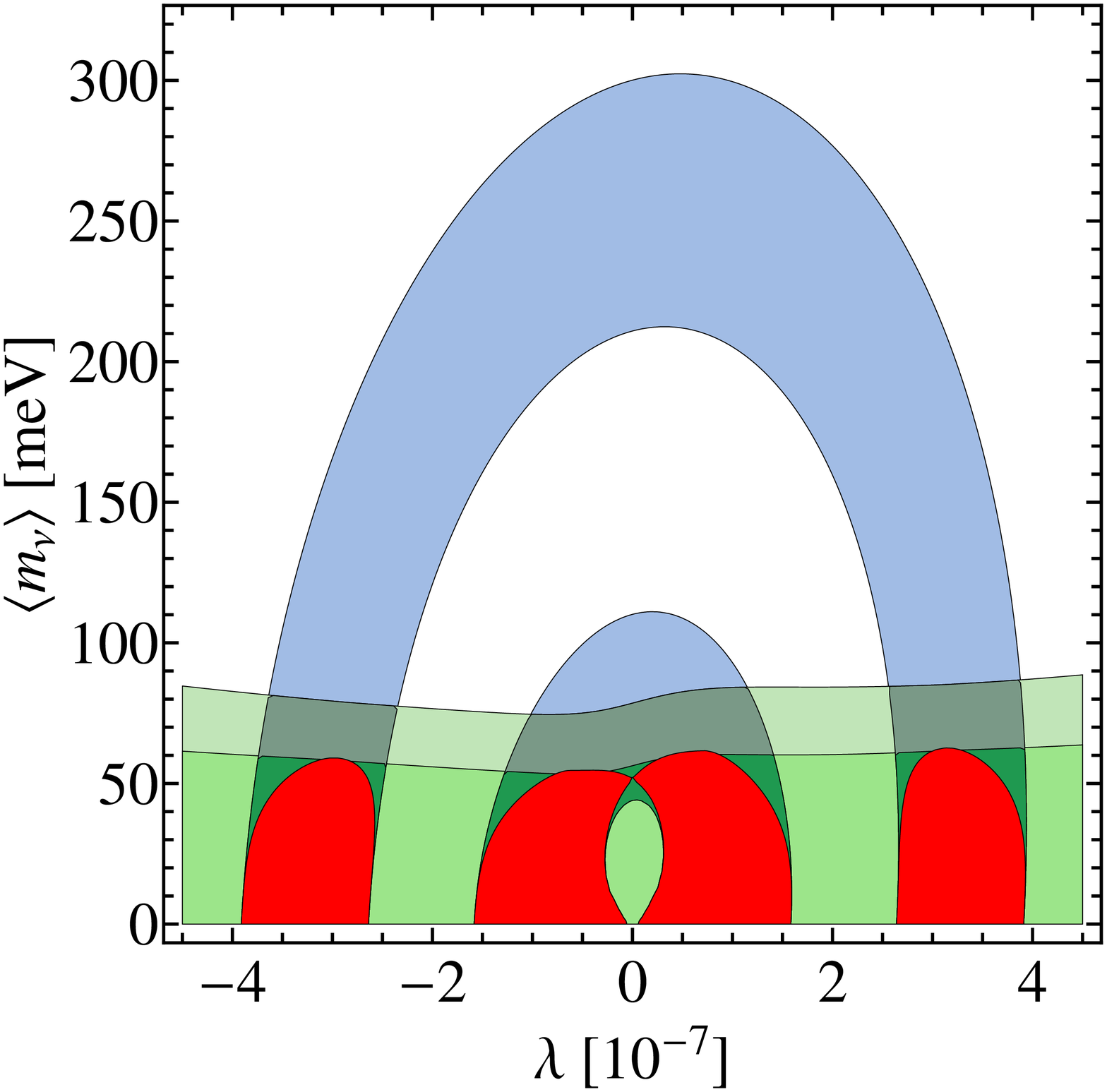}
}
\caption{Constraints at one standard deviation on the model parameters $m_\nu$ and $\lambda$ for $^{82}$Se from: (1) an observation of $0\nu\beta\beta$ decay half-life at $T_{1/2}=10^{25}$~y (outer blue elliptical contour) and $10^{26}$~y (inner blue elliptical contour); (2) reconstruction of the angular (outer, lighter green) and energy difference (inner, darker green) distribution shape; (3) combined analysis of (1) and (2) using decay rate and energy distribution shape reconstruction (red contours). The admixture of the MM and RHC$_\lambda$ contributions is assumed to be: (a) pure MM contribution; (b) 30\% RHC$_\lambda$ admixture; and (c) pure RHC$_\lambda$ contribution. NME uncertainties are assumed to be 30\% and experimental statistical uncertainties are determined from the simulation.}
\label{fig:discovery}
\end{figure*}

For the isotope $^{82}$Se, this is shown in Fig.~\ref{fig:discovery} for different RHC$_\lambda$ admixtures. The two blue elliptical contours correspond to the allowed one standard deviation ($m_\nu-\lambda$) parameter space at SuperNEMO when observing a signal at $T_{1/2} = 10^{25}$~y and $T_{1/2} = 10^{26}$~y, respectively. This takes into account a nominal theoretical uncertainty on the NME of 30\% and a one standard deviation statistical uncertainty on the measurement determined from the simulation (Fig.~\ref{fig:staterr}). The blue elliptical error bands therefore give the allowed parameter region when only considering the total $0\nu\beta\beta$ rate, which does not allow to distinguish between the MM and RHC$_\lambda$ contributions. 

When taking into account the information provided by the angular and energy difference distribution shape, the parameter space can be constrained significantly. This is shown using the green contours in Fig.~\ref{fig:discovery} for (a) a pure MM model, (b) a RHC$_\lambda$ admixture of 30\%, corresponding to an angular correlation of $k_{\theta}\approx 0.4$ and (c) a pure RHC$_\lambda$ model. This fixes two specific directions in the $m_\nu-\lambda$ plane (one for positive and one for negative $\lambda$). The widths of the contours are determined by the uncertainty in determining the theoretical correlation and admixture from the apparent distribution shape, see Fig.~\ref{fig:ktrue_krecon}. The outer (light green) contours in Fig.~\ref{fig:discovery} give the one standard deviation uncertainty on the parameters from reconstructing the angular distribution, while the inner (darker green) contour gives the one standard deviation uncertainty when using the distributions of the electron energy difference. As was outlined in Section~\ref{sec:simulation}, the energy difference distribution is expected to be easier to reconstruct and therefore gives a better determination of the RHC$_\lambda$ admixture and a better constraint. While interference between MM and RHC$_\lambda$ is neglected in the simulation, it is taken into account in Equation~(\ref{eq:T12_mu_lambda}) through the term $C_{m\lambda}\mu\lambda$ resulting in the slightly tilted elliptical contours and the asymmetry for $\lambda \leftrightarrow -\lambda$.
\begin{figure*}[!t]
\centering
\subfloat[][]{
\includegraphics[clip,width=0.3\textwidth]{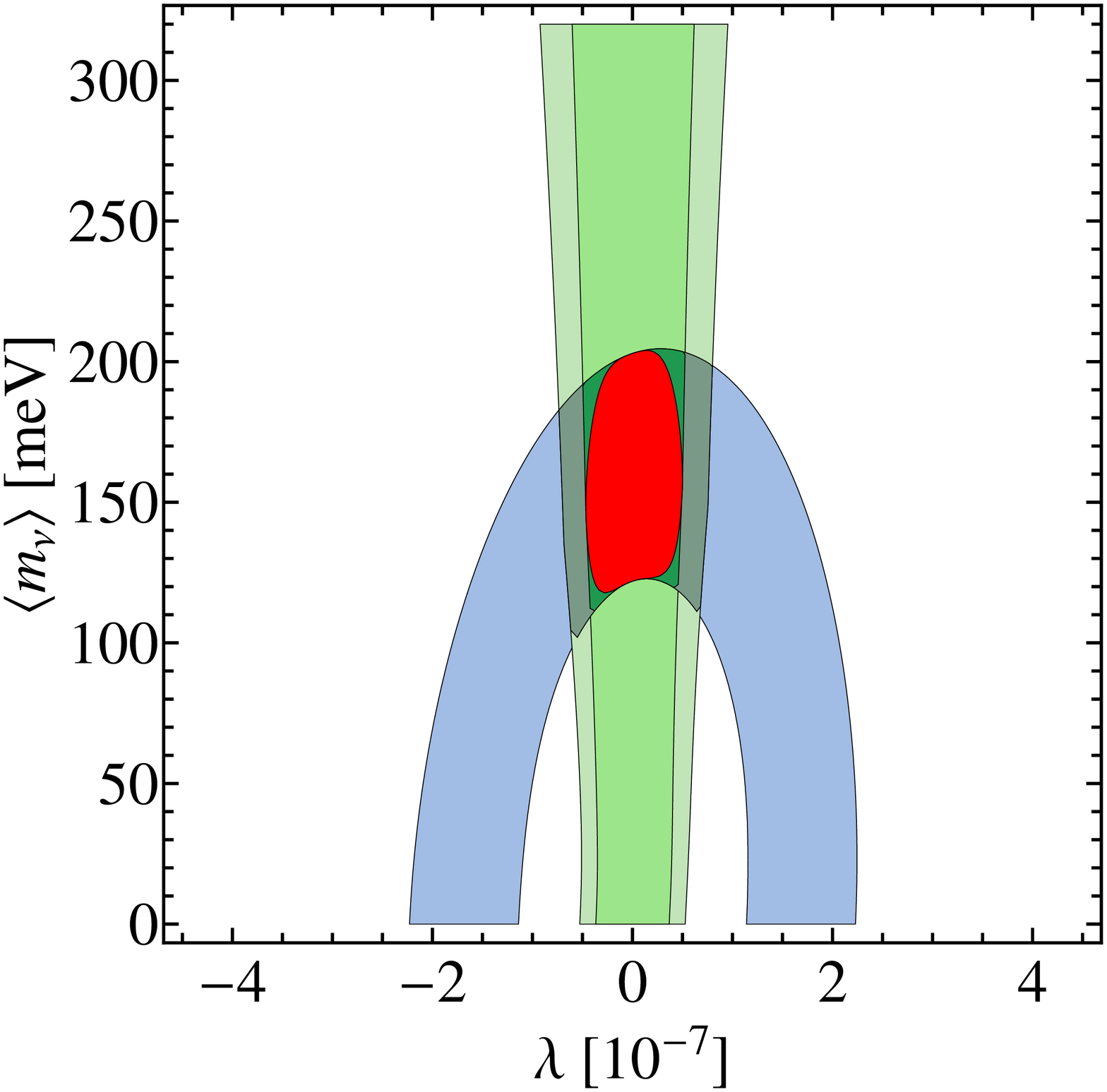}
}
\subfloat[][]{
\includegraphics[clip,width=0.3\textwidth]{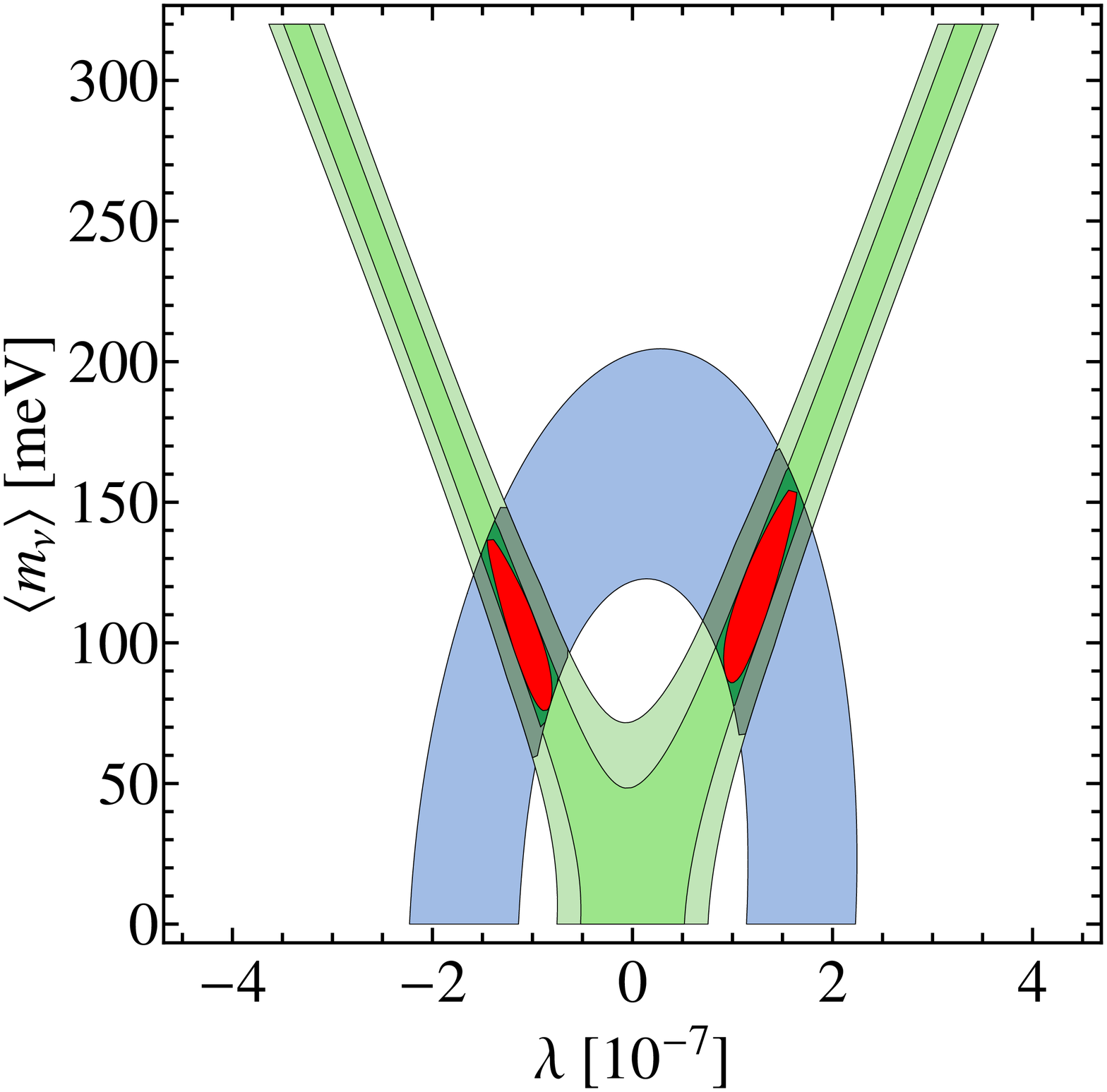}
}
\subfloat[][]{
\includegraphics[clip,width=0.3\textwidth]{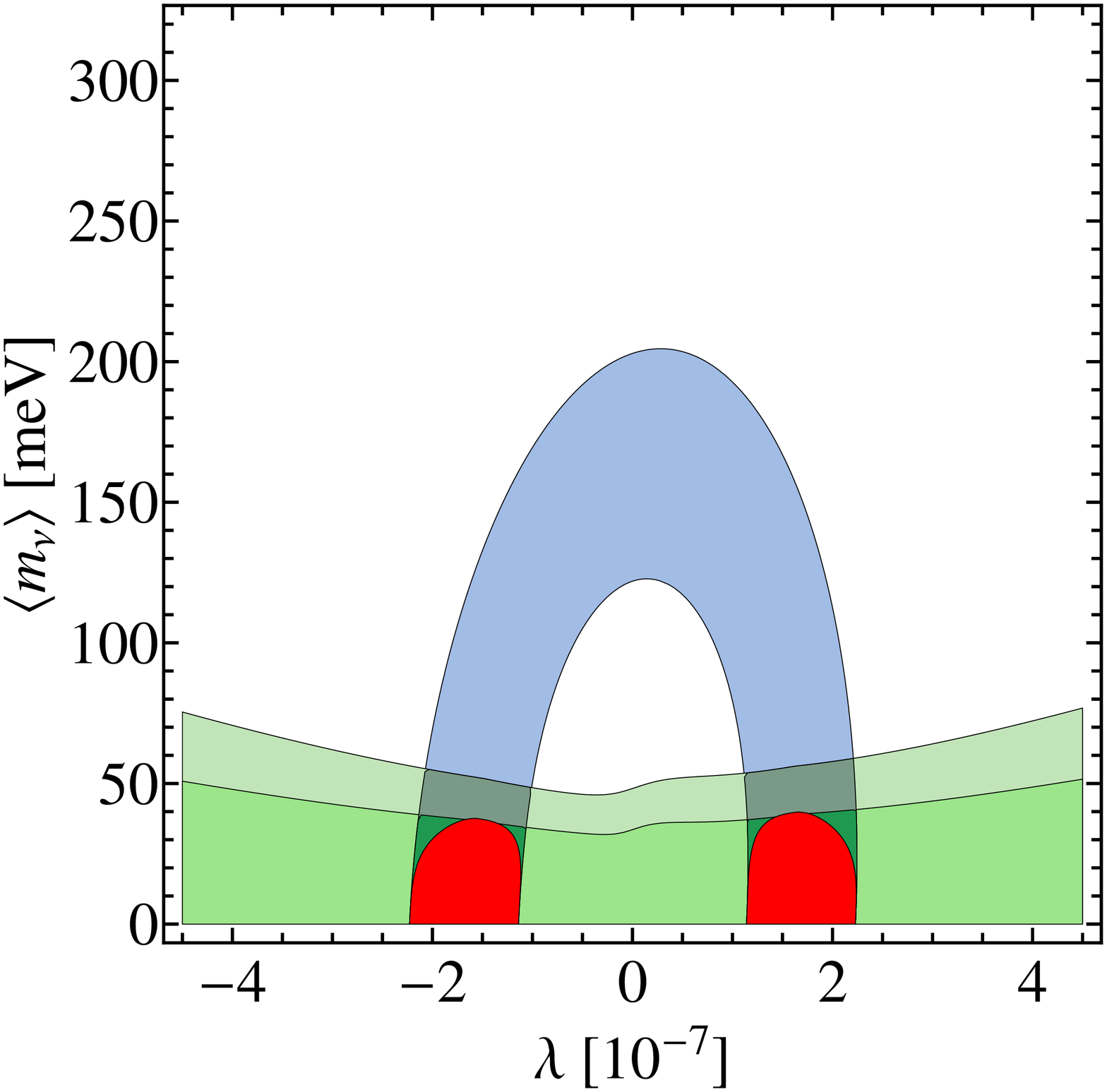}
}
\caption{As Fig.~\ref{fig:discovery} but for the isotope $^{150}$Nd with a decay half-life of $T_{1/2}=10^{25}$~y.}
\label{fig:discoveryNd}
\end{figure*}
Finally, the red contours in Fig.~\ref{fig:discovery} show the constraints on the model parameters when combining both the determination of the $0\nu\beta\beta$ decay rate and the decay energy distribution. This demonstrates that such a successful combination can make it possible to determine the mechanism (i.e. the degree of MM and RHC$_\lambda$ admixture in this case), and provide a better constraint on the model parameters. From Fig.~\ref{fig:discovery} (a), the Majorana mass term can be determined at $\langle m_\nu \rangle=245^{+56}_{-41}$ meV while the $\lambda$ parameter is constrained to be $-0.87\cdot10^{-7} < \lambda < 0.92\cdot10^{-7}$ in the case of a measured $0\nu\beta\beta$ decay half-life of $^{82}$Se of $T_{1/2} = 10^{25}$~y. For a $^{82}$Se half-life of $T_{1/2} = 10^{26}$~y, the uncertainty on the decay rate increases as SuperNEMO reaches its exclusion limit for RHC$_\lambda$ admixtures. It is therefore only possible to extract upper limits on the model parameters from Fig.~\ref{fig:discovery} for $T_{1/2} = 10^{26}$~y. However, the shape information provides additional constraints on the parameter space. In Fig.~\ref{fig:discoveryNd} we show the analogous plots for the isotope $^{150}$Nd assuming a decay half-life of $T_{1/2}=10^{25}$~y.

%------------------------------------------------------------------------------
\subsection{Rate Comparison of $^{\bf{150}}$Nd and $^{\bf{82}}$Se}\label{sec:5050}

While reconstruction of the decay distribution can be an ideal way to distinguish between different mechanisms, it might be of little help if $0\nu\beta\beta$ decay is observed close to the exclusion limit of SuperNEMO, or not at all. This is demonstrated in Fig.~\ref{fig:discovery} where, for a half-life of $T_{1/2}=10^{26}$~y, the reconstruction of the energy difference distribution will be problematic due to the low number of events (compare Fig.~\ref{fig:ktrue_krecon}). As an alternative, it is possible to compare the $0\nu\beta\beta$ rate in different isotopes. This method, which could provide crucial information close to the exclusion limit, is especially relevant for SuperNEMO which could potentially measure $0\nu\beta\beta$ decay in two (or more) isotopes. Such a comparative analysis was used in~\cite{Deppisch:2006hb} to distinguish between several new physics mechanisms. A combined analysis of several isotopes, potentially measured in other experiments, will improve the statistical significance~\cite{Gehman:2007qg}.

The possibility of sharing the two isotopes equally in SuperNEMO, each with a total exposure of 250~kg~y, is now considered. In the cases where the MM or the RHC$_\lambda$ contributions dominate, the following half-life ratios can be found:
\begin{eqnarray}
	\label{mmratio}
	&\mathrm{MM}:& \quad
	\frac{T_{1/2}^{^{82}\mathrm{Se}}}{T_{1/2}^{^{150}\mathrm{Nd}}} =
        \frac{C_{mm}^{^{150}\mathrm{Nd}}}{(2.7)^2 \cdot C_{mm}^{^{82}\mathrm{Se}}} =
        2.45, \\
	\label{RHCratio}
	&\mathrm{RHC}_\lambda:& \quad
	\frac{T_{1/2}^{^{82}\mathrm{Se}}}{T_{1/2}^{^{150}\mathrm{Nd}}} =
    \frac{C_{\lambda\lambda}^{^{150}\mathrm{Nd}}}{(2.7)^2\cdot C_{\lambda\lambda}^{^{82}\mathrm{Se}}} =
        3.64.
\end{eqnarray}
These ratios and their uncertainties are determined by the $0\nu\beta\beta$ decay NMEs and phase spaces. The factor 2.7 is the correction added to the $^{150}$Nd NMEs as described in Section~\ref{sec:nme}. It has recently been suggested that uncertainties in NME calculations are highly correlated~\cite{correlate} so measurements made with two or more isotopes could reduce the uncertainty on the physics parameters significantly. Additionally, most experimental systematic uncertainties would cancel if different isotopes are studied in a single experiment such as SuperNEMO. This would not be possible when comparing results with other experiments. The statistical uncertainties are naturally greater than in the single-isotope case, due to the exposure being halved for each isotope. 

\begin{figure*}[!t]
\centering
\subfloat[][]{
\includegraphics[clip,width=0.3\textwidth]{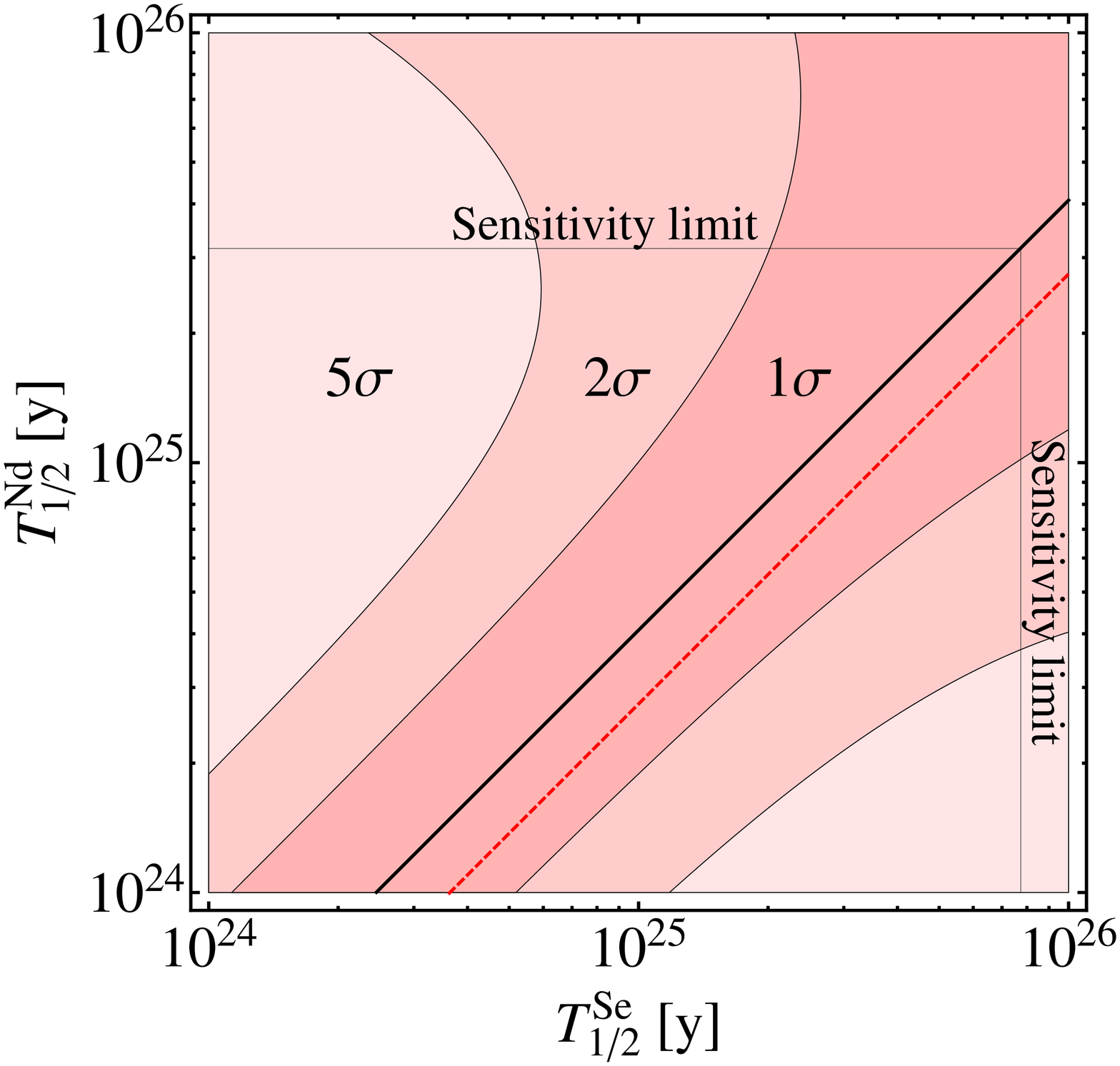}
}
\subfloat[][]{
\includegraphics[clip,width=0.3\textwidth]{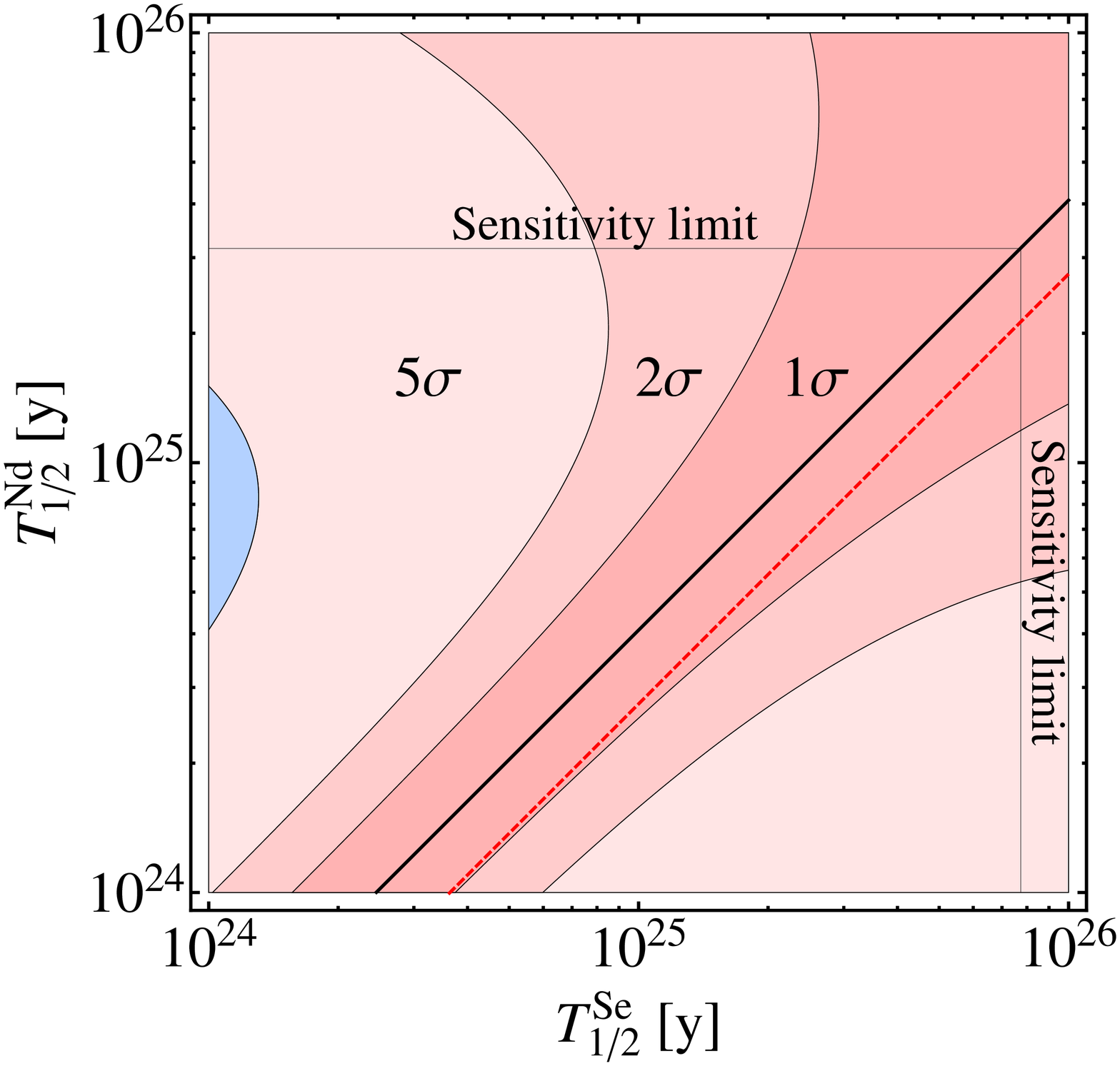}
}
\subfloat[][]{
\includegraphics[clip,width=0.3\textwidth]{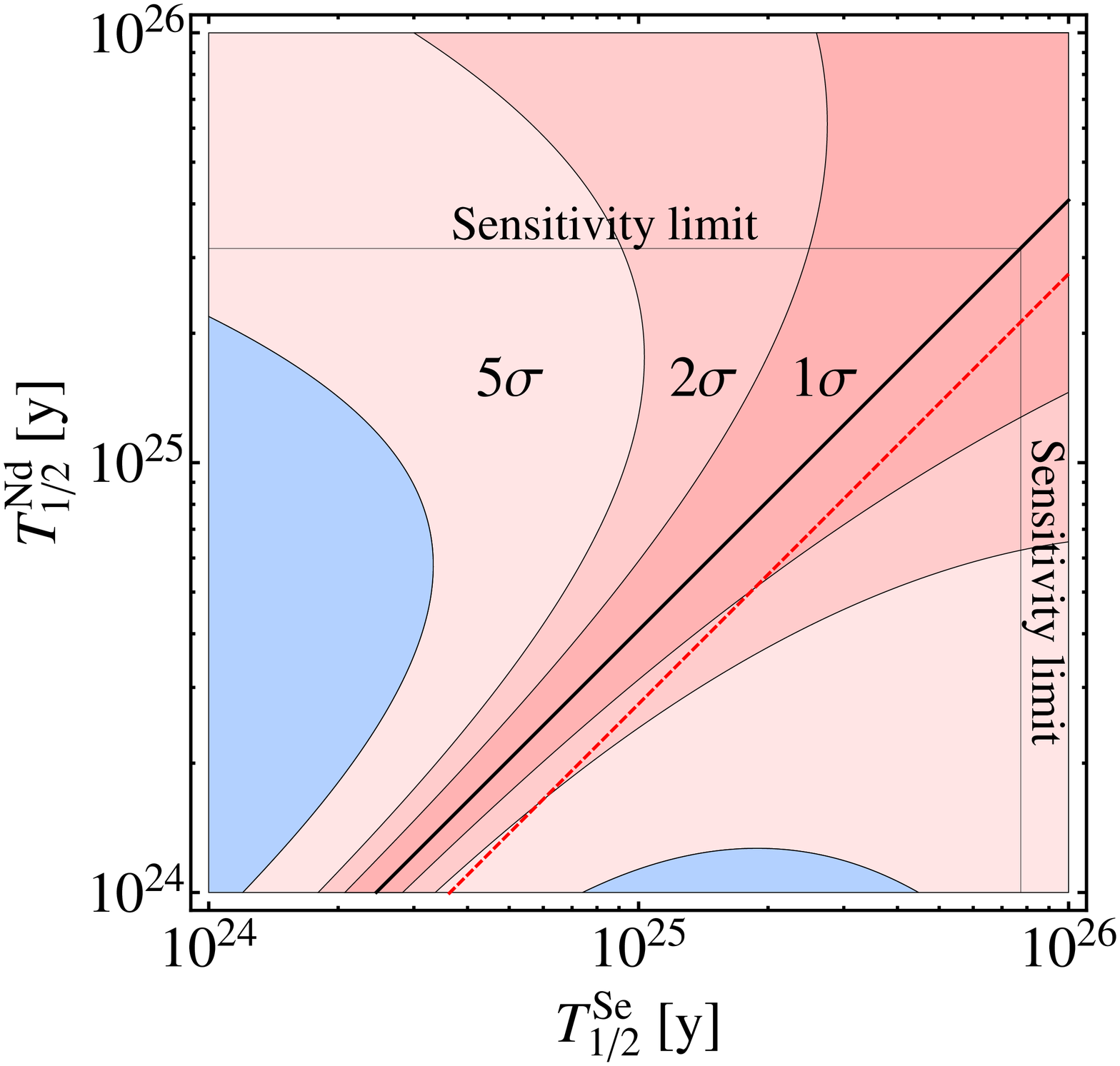}
}
\caption{The $0\nu\beta\beta$ half-life of $^{150}$Nd as a function of measured half-life in $^{82}$Se for the hypothesis that the MM is the single $0\nu\beta\beta$ decay source. The contours show the 1, 2 and 5 standard deviation levels assuming statistical uncertainties derived from the experimental simulation and 30\% NME errors assumed to have (a) no, (b) 0.7 and (c) perfect correlation.  The experimental uncertainties and expected sensitivity ($90\%$ CL exclusion) limit are calculated for 250~kg~y of exposure (assuming a 50\% $^{82}$Se and 50\% $^{150}$Nd option). The red line shows the relationship for the RHC$_\lambda$. The blue contour shows the 5$\sigma$ exclusion of the MM.}
\label{fig:5050compare}
\end{figure*}

The results of the combined NME and statistical uncertainties analysis, including a possible correlation of the NMEs, are illustrated in Fig.~\ref{fig:5050compare}. It shows the $0\nu\beta\beta$ half-life of $^{150}$Nd as a function of the half-life in $^{82}$Se assuming a pure MM model, with the coloured contours giving the deviation from the hypothesis that the mass mechanism is the single source of $0\nu\beta\beta$ decay in both isotopes at the 1, 2 and 5 standard deviation level.  The statistical uncertainties used in Fig.~\ref{fig:5050compare} are derived from our experimental simulation and the standard 30\% NME uncertainties are applied. The effect of a possible correlation of the NMEs is shown by assuming the covariance coefficient between the NME uncertainties of $^{82}$Se and $^{150}$Nd to be (a) zero (no correlation), (b) 0.7 and (c) 1.0 (full correlation).  The experimental uncertainties and expected sensitivity ($90\%$ CL exclusion) limits are calculated for 250~kg~y of exposure of each isotope and assume a 50\% $^{82}$Se and 50\% $^{150}$Nd option for SuperNEMO. The red line shows the relationship for the half-life ratio in the pure RHC$_\lambda$ model (Equation~(\ref{RHCratio})). It can be seen that an exclusion at two standard deviations is possible if the NME errors are perfectly correlated and at the one standard deviation level if the correlation is 70\%, which is a more realistic assumption. 

Other mechanisms have different half-life ratios~\cite{Deppisch:2006hb} so they could be excluded with different CLs at SuperNEMO. One important advantage of this method is that it provides a possibility to falsify the mass mechanism as the sole source for $0\nu\beta\beta$. A measurement within the blue contour would indicate that the pure MM model can be excluded at the 5 standard deviation level and new physics is required to explain the measured half-lives.

%------------------------------------------------------------------------------
\subsection{Combined Energy and Rate Comparison of $^{\bf{150}}$Nd and $^{\bf{82}}$Se}\label{sec:AE5050}

\begin{figure*}[!t]
\centering
\subfloat[][]{
\includegraphics[clip,width=0.3\textwidth]{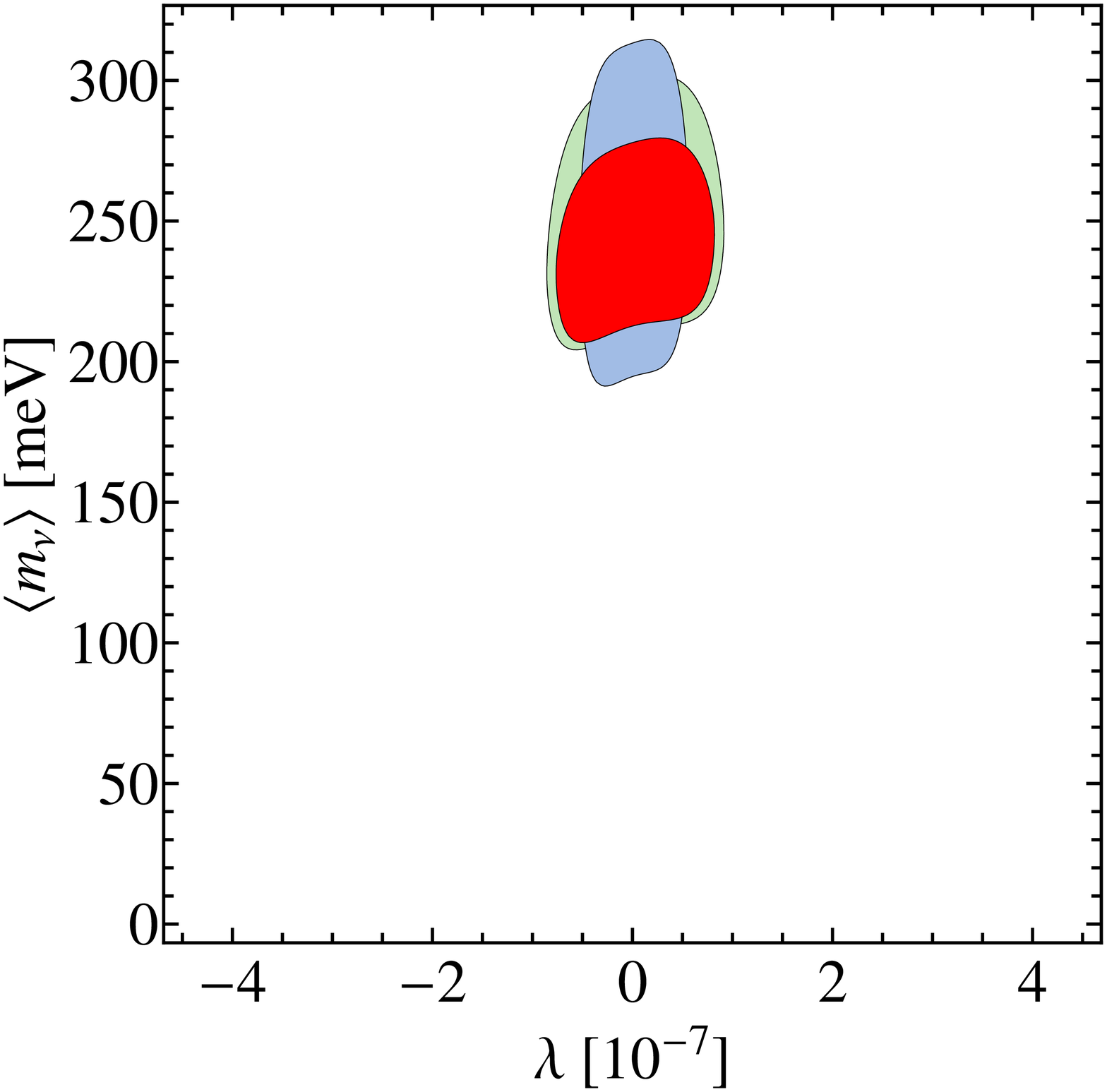}
}
\subfloat[][]{
\includegraphics[clip,width=0.3\textwidth]{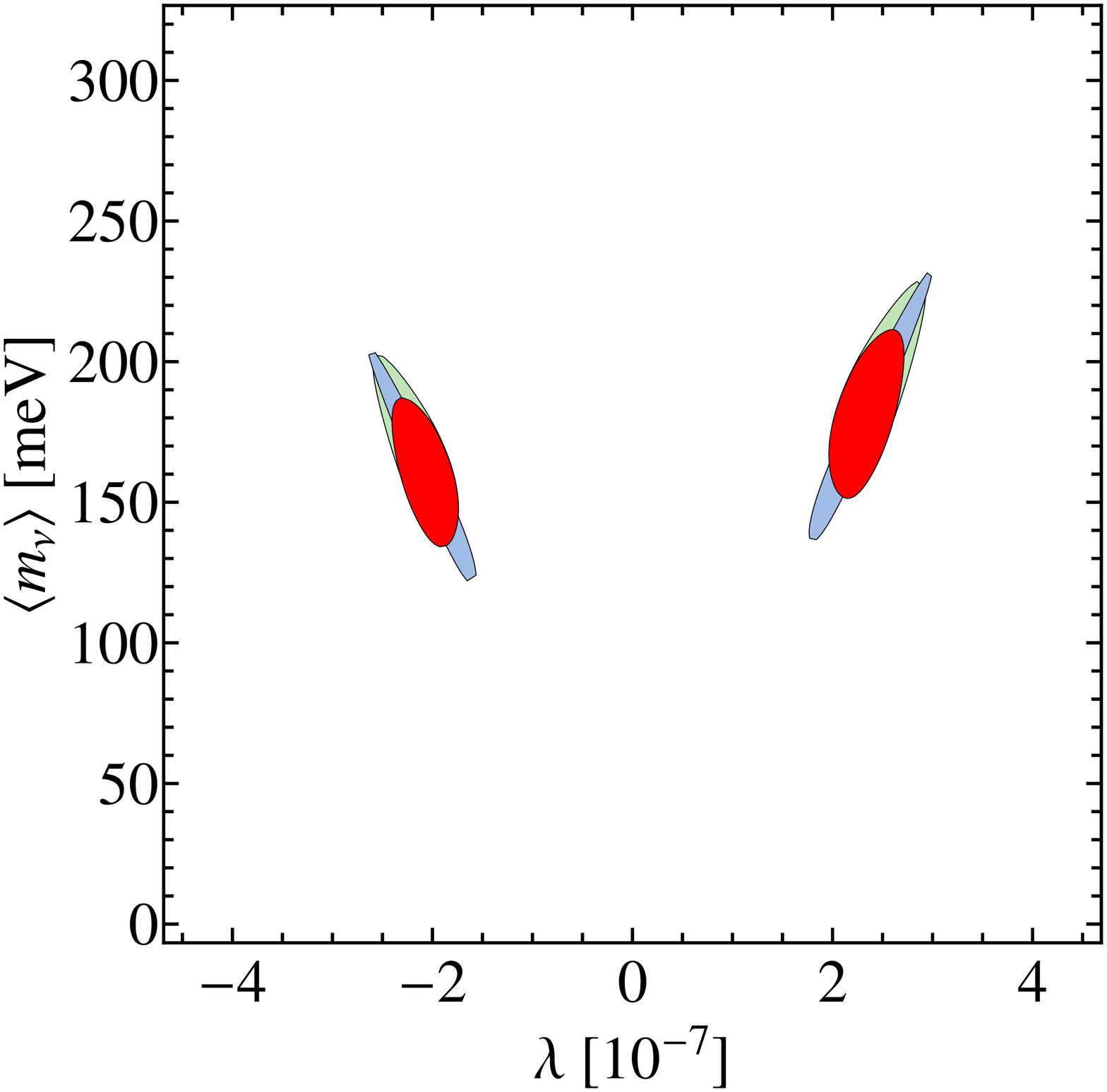}
}
\subfloat[][]{
\includegraphics[clip,width=0.3\textwidth]{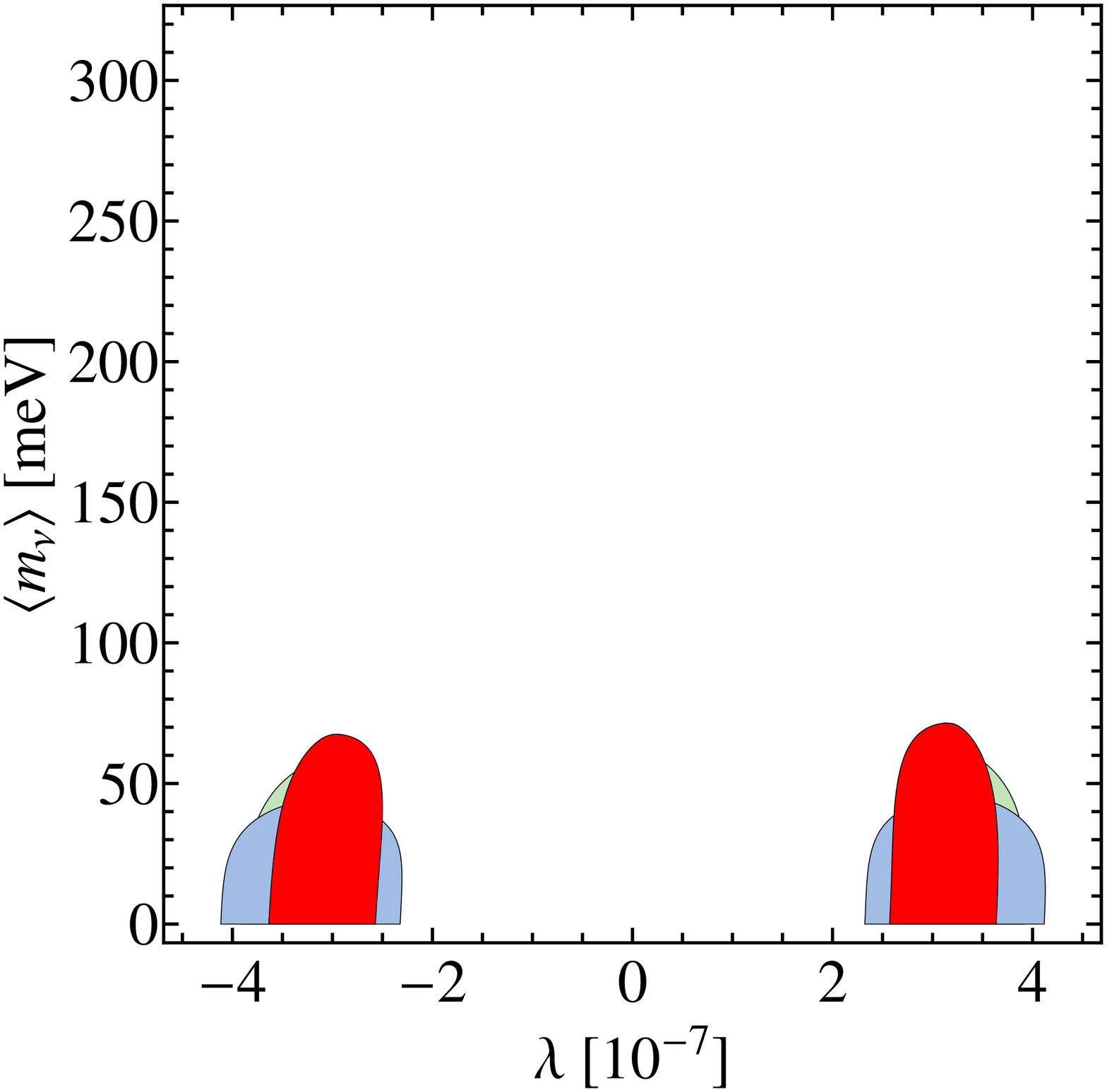}
}
\caption{Constraints at one standard deviation on the model parameters $m_\nu$ and $\lambda$ from: (1) an observation of $0\nu\beta\beta$ decay half-life of $^{82}$Se at $T_{1/2}=10^{25}$~y with 500 kg y exposure and reconstruction of the energy difference distribution (outer green contour); (2) an observation of $0\nu\beta\beta$ decay half-life of $^{150}$Nd at a half-life corresponding to $T_{1/2}=10^{25}$~y in $^{82}$Se with an exposure of 500 kg y and reconstruction of the energy difference distribution (inner blue contour); (3) combined analysis of (1) and (2) with an exposure of 250 kg y in $^{82}$Se and $^{150}$Nd (red contour). The admixture of the MM and RHC$_\lambda$ contributions is assumed to be: (a) pure MM contribution; (b) $30\%$ RHC$_\lambda$ admixture; and (c) pure RHC$_\lambda$ contribution. NME uncertainties are assumed to be 30\% with a correlation of the uncertainties of 0.7, and experimental statistical uncertainties are determined from the simulation.}
\label{fig:discoveryCombo}
\end{figure*}

In the most favourable case, signal event rates in two isotopes could be high enough ($0\nu\beta\beta$ decay half-lives small enough) that the distribution method and the two isotope rate analysis can be combined to put further constraints on the parameter space. The effect of such a combined analysis on the allowed parameter space is shown in Fig.~\ref{fig:discoveryCombo}, where the 50\% $^{150}$Nd - 50\% $^{82}$Se two-isotope option (red contours) is compared to the single-isotope options 100\% $^{82}$Se (green contours) and 100\% $^{150}$Nd (blue contours). The $0\nu\beta\beta$ decay half-life of $^{82}$Se is assumed to be $10^{25}$~y, and the half-life of $^{150}$Nd is determined by the respective MM-RHC$_\lambda$ admixture, i.e. (a) $T_{1/2}^{Nd}=10^{25}/2.45$~y, (b) $10^{25}/2.73$~y and (c) $10^{25}/3.64$~y. The NME uncertainties are assumed to be 30\% with a 0.7 covariance between the uncertainties of the NMEs of $^{82}$Se and $^{150}$Nd. As can be seen in Fig.~\ref{fig:discoveryCombo}, the two-isotope option can improve the constraints on the parameter space along the radial direction, e.g. it allows a more accurate determination of the MM neutrino mass $m_\nu$ in Fig.~\ref{fig:discoveryCombo} (a). On the other hand, the accuracy in the lateral direction (the parameter $\lambda$ in Fig.~\ref{fig:discoveryCombo} (a)) becomes worse compared to the best single-isotope option due to the reduced statistics for a given isotope.

%

%------------------------------------------------------------------------------
\section{Conclusion}\label{sec:conclusion}

The $0\nu\beta\beta$ decay is a crucial process for physics beyond the Standard Model, and the next generation SuperNEMO experiment is designed to be a sensitive probe of this lepton number violating observable. In addition to being able to measure the $0\nu\beta\beta$ half-life of one or more isotopes, it also allows the determination of the angular and energy difference distributions of the outgoing electrons.

In this paper we have focussed on the sensitivity of SuperNEMO to new physics and its ability to discriminate between different $0\nu\beta\beta$ mechanisms. This was achieved by a detailed analysis of two important models, namely the standard mass mechanism via light left-handed Majorana neutrino exchange and a contribution from right-handed current via the effective $\lambda$ parameter stemming from Left-Right symmetry. The study included a full simulation of the process and the SuperNEMO detector at the event level, allowing a realistic estimation of the experimental $90\%$~CL exclusion limit and statistical uncertainties.

SuperNEMO is expected to exclude $0\nu\beta\beta$ half-lives up to $1.2\cdot10^{26}$~y (MM) and $6.1\cdot10^{25}$~y (RHC$_\lambda$) for $^{82}$Se and $5.1\cdot 10^{25}$~y (MM) and $2.6\cdot10^{25}$~y (RHC$_\lambda$) for $^{150}$Nd at $90\%$ CL with a detector exposure of 500 kg y. This corresponds to a Majorana neutrino mass of $m_\nu\approx 70$~meV and a $\lambda$ parameter of $\lambda\approx 10^{-7}$, giving an improvement of more than one order of magnitude compared to the NEMO-III experiment.

It has been shown that the angular and electron energy difference distributions can be used to discriminate new physics scenarios. In the framework of the two mechanisms analysed, it was demonstrated that using this technique the individual new physics model parameters can be determined. For a half-life of $T_{1/2}=10^{25}$~y with an exposure of 500~kg~y, the Majorana neutrino mass can be determined to be 245 meV with an uncertainty of 30\% while the $\lambda$ parameter can be constrained at the same time to be smaller than $|\lambda|<0.9\cdot10^{-7}$. Such a decay distribution analysis could be easily extended further to include other new physics scenarios with distinct distributions and the results are quoted in terms of a generalised distribution asymmetry parameter to allow new physics scenarios to be compared. As the two example mechanisms considered exhibit maximally different angular and energy distribution shapes, they serve as representative scenarios covering a large spectrum of the model space. For example, the right-handed current contribution due the effective $\eta$ parameter, also arising in Left-Right symmetrical models, can be distinguished from the mass mechanism and the right-handed current $\lambda$ contribution by looking at both the angular and energy difference decay distribution. This would allow a determination of all three model parameters $m_\nu$, $\lambda$ and $\eta$ by looking at the total rate and the angular and energy difference distribution shapes.

Further insight into the mechanism of $0\nu\beta\beta$ can be gained by using multiple isotopes within the SuperNEMO setup. This possibility was explored by studying the option of having 50$\%$ $^{150}$Nd and 50$\%$ $^{82}$Se, each with an exposure of 250~kg~y. While the statistics per isotope is reduced compared to the individual 100\% options, the ability to measure the ratio between the half-lives of the two isotopes can be used as additional information on the underlying physics mechanism responsible for $0\nu\beta\beta$ decay. As was shown for the isotopes $^{82}$Se and $^{150}$Nd at SuperNEMO, this could be a powerful method to falsify the mass mechanism as the dominant $0\nu\beta\beta$ mechanism. A correlation between the uncertainties of nuclear matrix elements, which is generally expected on theoretical grounds, has proven to be of importance and its impact on the falsification potential was analysed. Within SuperNEMO such a correlation could also be found between the systematic uncertainties in the measurements of different isotopes.

SuperNEMO also has a number of other possibilities to disentangle the underlying physics. The detection technology is not dependent on one particular isotope and any double $\beta$ emitting source could be studied in the detector. In this paper $^{82}$Se and $^{150}$Nd have been considered but other isotopes such as $^{48}$Ca or $^{100}$Mo are feasible. The analysis can be extended to cover more than two isotopes thereby achieving a higher significance and a comparison with other experimental results will provide additional information. SuperNEMO is also able to measure a $0\nu\beta\beta$ decay to an excited state, by measuring two electrons and an accompanying photon. This again could be used to aid the analysis to discriminate between new physics mechanisms. 

A combination of the above methods makes SuperNEMO an exciting test of new physics. These methods would prove invaluable in excluding or confirming dominating mechanisms of lepton number violation in the reach of the next generation $0\nu\beta\beta$ experiments.

%------------------------------------------------------------------------------
\begin{acknowledgement}
The authors would like to thank H.~P\"as, M.~Hirsch, E.~Lisi, V.~Rodin and A.~Faessler for useful discussions. We acknowledge support by the Grants Agencies of France, the Czech Republic, RFBR (Russia), STFC (UK), MICINN (Spain), NSF, DOE, and DOD (USA). We acknowledge technical support from the staff of the Modane Underground Laboratory (LSM).
\end{acknowledgement}
%-----------------------------------------------------------------------------
%
% BibTeX users please use
% \bibliographystyle{}
% \bibliography{}

\end{document}